\documentclass{article}

\usepackage{amsmath}
\usepackage{mathtools}
\usepackage{tikz}
\usepackage{mathdots}
\usepackage{amsfonts}
\usepackage{amsthm}
\usepackage{latexsym}   
\usepackage{graphicx}
\usepackage{color}
\usepackage{hyperref}
\usepackage[linesnumbered,ruled,vlined]{algorithm2e}
\usepackage{booktabs}
\usepackage{arydshln}
\usepackage{bm}
\usepackage{multirow}
\usepackage{makecell}
\usepackage{booktabs}
\usepackage{balance}
\usepackage{subcaption}
\usepackage{fullpage}
\usepackage{authblk}
\usepackage[numbers,sort&compress]{natbib}
\usepackage{amssymb}
\usepackage[T1]{fontenc}

\RequirePackage[edges]{forest}
\usetikzlibrary{arrows.meta}
\usetikzlibrary{tikzmark}
\usetikzlibrary{bending}
\RequirePackage{multicol}
\usetikzlibrary{shadows}
\RequirePackage{tikz-qtree}
\usetikzlibrary{shadows.blur}

\usepackage{xcolor}

\newtheorem{theorem}{Theorem}[section]

\theoremstyle{definition}

\theoremstyle{remark}

\theoremstyle{plain}
\newtheorem{problem}[theorem]{Problem}       

\AtBeginDocument{%
  }

\usepackage[many]{tcolorbox}
\usepackage{xcolor}

\title{A Survey on Algorithmic Interventions in Opinion Dynamics}

\author[1]{Atsushi Miyauchi\thanks{atsushi.miyauchi@intesasanpaolo.com}}
\author[2]{Yuko Kuroki\thanks{yuko.miyauchi@intesasanpaolo.com}}
\author[2]{Federico Cinus\thanks{federico.cinus@intesasanpaolo.com}}
\author[3]{Stefan Neumann\thanks{stefan.neumann@tuwien.ac.at}}
\author[2]{Francesco~Bonchi\thanks{francesco.bonchi@intesasanpaolo.com}}

\affil[1]{Intesa Sanpaolo, Turin, Italy}
\affil[2]{Intesa Sanpaolo AI Research, Turin, Italy}
\affil[3]{TU Wien, Vienna, Austria}

\begin{document}

\maketitle

\begin{abstract} 
Social media platforms have become critical infrastructures for public communication, where large-scale interaction can both support socially beneficial collective pressure and amplify polarization and conflict. While opinion-dynamics research has long modeled how beliefs evolve through interpersonal influence, the central challenge for healthier online environments increasingly lies in algorithmic interventions: mechanisms that steer collective opinion toward desirable outcomes or dampen harmful dynamics. This survey offers a structured synthesis of this fast-growing, interdisciplinary literature. We organize prior work by the objective optimized---overall opinion (e.g., consensus or mean opinion), polarization and disagreement, and other quantities---and review the associated optimization formulations and representative algorithms with mathematical rigor. We also compile intervention-relevant theoretical and empirical findings. Finally, we outline concrete future directions that emerge from this survey. 
\end{abstract}

\section{Introduction}\label{sec:intro}

Social media platforms such as $\mathbb{X}$ and Facebook increasingly function as critical infrastructures for public communication, shaping the space in which users debate and public opinion coalesces. Although these opinion-formation processes can benefit society---for example, by generating collective pressure against undesirable political narratives~\cite{Gainous+13,Van+13}---they may also amplify polarization and conflict, thereby exacerbating societal divisions and potentially undermining social stability~\cite{Bail22}.

The study of opinion dynamics predates social media: researchers have long investigated how individual beliefs evolve through interpersonal influence. With the rise of social media, opinion dynamics have become dramatically more consequential in practice, as influence can propagate at unprecedented scale and speed and rapidly shape collective outcomes. At the same time, the availability of large-scale, high-resolution behavioral data and new interaction mechanisms have further accelerated this line of work. A central aim of this literature is to develop models that capture key qualitative and quantitative features of opinion formation in society.

However, understanding opinion dynamics alone is often insufficient for promoting healthier online environments. Achieving this goal also requires effective \emph{algorithmic interventions}---approaches that can steer collective opinions toward socially desirable outcomes (e.g., mobilizing public opposition to undesirable political narratives) or mitigate harmful dynamics (e.g., polarization and conflict). A growing body of research addresses such interventions; however, it spans a wide range of objectives and relies on diverse opinion-dynamics models in different settings, making it difficult to systematically synthesize prior work and assess the state of the art. Moreover, this intervention-focused research is inherently interdisciplinary, spanning computer science as well as control theory and statistical physics.

\smallskip
\noindent
\textbf{Survey Contributions and Roadmap.}
This survey provides a structured overview of the rapidly growing literature on algorithmic interventions in opinion dynamics. We organize the review around the \emph{objective} of the intervention: Section~\ref{sec:overall} covers work on optimizing overall opinion in the network, Section~\ref{sec:pd} focuses on polarization and disagreement, and Section~\ref{sec:others} reviews other objectives. For each category, we examine the underlying optimization formulations and representative algorithms across different intervention mechanisms, 
covering both classic and recent results in a mathematically grounded manner.
Beyond explicit intervention models and algorithms, we also summarize intervention-relevant insights, namely studies that do not directly propose an intervention mechanism in opinion dynamics but offer theoretical or empirical guidance for intervention design.
Figure~\ref{fig:map} provides a bird's-eye view of our taxonomy. Each of the three technical sections (Sections~\ref{sec:overall}--\ref{sec:others}) is accompanied by a table (Tables~\ref{tab:overall_summary}--\ref{tab:summary_others}) summarizing the main papers along several key dimensions, such as the objective, the type of underlying graph, and the opinion-dynamics model.
Before turning to the technical sections, Section~\ref{sec:models} briefly reviews the opinion-dynamics models most commonly used in studies of algorithmic interventions, including the DeGroot and Friedkin--Johnsen models. Finally, Section~\ref{sec:conclusion} concludes the survey and outlines directions for future research.

\smallskip
\noindent
\textbf{Related Work.}
Although several comprehensive surveys on opinion-dynamics models have recently become available~\cite{Peralta+25,Starnini+25,Shirzadi+25,Zareer+25_2}, they do not detail algorithmic interventions. The surveys most closely related to ours are those by \citet{Li+21}, \citet{hartman2022interventions}, and \citet{Interian+23}, which review algorithmic interventions aimed at controlling quantities related to societal divisions and social stability (e.g., polarization and conflict). 
Our survey differs from these in two key respects: (i) we cover a broader range of quantities to optimize beyond societal-division and stability-related quantities, including overall opinion (e.g., consensus or mean opinion), for which a substantial body of work exists; and (ii) the most recent of these surveys~\cite{Interian+23} was published in 2023, whereas the literature has grown rapidly since then, with a number of important papers appearing even within the narrower scope of controlling polarization and conflict. 

We acknowledge the vast literature on \emph{influence maximization}---typically, selecting seed nodes (and sometimes modifying network structure) to maximize the spread of an information-diffusion process~\cite{Kempe+03,Kempe+15}. 
However, influence maximization is out of scope for this survey. Our focus is on opinion dynamics in society as opposed to simple cascade models of information diffusion. In many settings, opinion dynamics is better modeled as a sequence of interpersonal interactions that iteratively update individuals' beliefs and may converge to an equilibrium, rather than as a one-shot cascading adoption process. Moreover, opinions are typically not binary states (e.g., active vs. inactive); instead, they are often represented as continuous values capturing the strength or valence of an individual's attitude toward a topic.

\definecolor{myblue}{HTML}{DAE8FC}
\definecolor{mypink}{HTML}{F8CECC}
\definecolor{mygrey}{HTML}{F5F5F5}
\definecolor{myborder}{HTML}{6A879E}

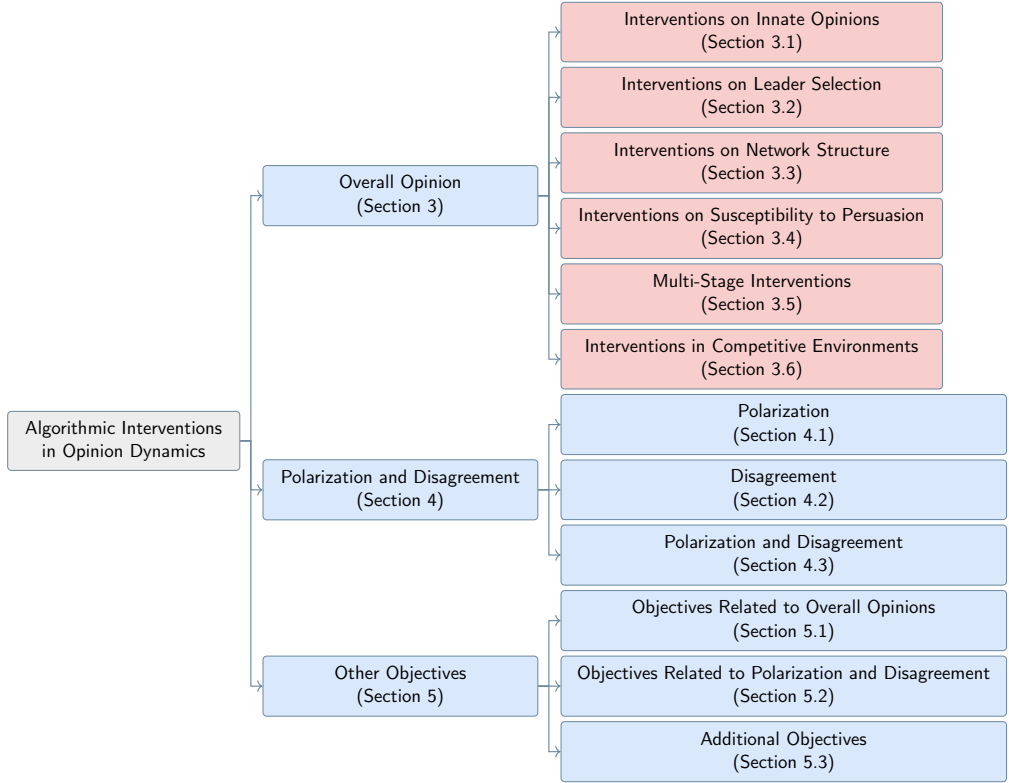
\begin{figure}[t]
    \centering
    \resizebox{0.82\textwidth}{!}{
    \tikzset{
        basic/.style  = {
            draw=myborder,
            align=center, 
            font=\sffamily, 
            rectangle,         
            inner sep=1.4pt
        },
        root/.style = {
            basic, 
            rounded corners=2pt, 
            thin, 
            align=center, 
            inner sep=5pt,
            fill=black!7, 
            text width=4.0cm,
            anchor=west 
        },
        obj/.style = {
            basic, thin, rounded corners=2pt, fill=myblue,
            align=center, 
            text width=4.8cm, 
            inner sep=5pt,
            anchor=west, 
            child anchor=north,
            parent anchor=south,
      },
      obj_det/.style = {
            basic, thin, rounded corners=2pt, fill=myblue,
            align=center, 
            text width=8.0cm, 
            inner sep=5pt,
            anchor=west, 
            child anchor=north,
            parent anchor=south,
      },
       tnode/.style  = {basic, thin, rounded corners=2pt, align=center, inner sep=5pt, fill=mypink, text width=6.8cm},
       onode/.style  = {basic, thin, rounded corners=2pt, align=left, fill=mygrey, 
       text width=6cm
       },
    }
    \begin{forest}
        for tree={
            s sep=2.5pt,
            grow'=0,
            anchor=west,
            child anchor=west,
            tier/.option=level, 
            forked edges,         
            fork sep=2mm,              
            edge={semithick, ->, draw=myborder},   
        }
        [Algorithmic Interventions\\ in Opinion Dynamics, root,
        calign=center, 
        parent anchor=east,  
            [Overall Opinion\\(Section~\ref{sec:overall}), obj,
            child anchor=west,
            parent anchor=east,
                [Interventions on Innate Opinions\\ (Section~\ref{subsec:overall_innate}), tnode
                ]
                [Interventions on Leader Selection\\ (Section~\ref{subsec:overall_leader}), tnode
                ]
                [Interventions on Network Structure\\ (Section~\ref{subsec:overall_structure}), tnode
                ]
                [Interventions on Susceptibility to Persuasion\\ (Section~\ref{subsec:overall_susceptibility}), tnode
                ]
                [Multi-Stage Interventions\\
                (Section~\ref{subsec:overall_multi}), tnode
                ]
                [Interventions in Competitive Environments\\ (Section~\ref{subsec:overall_competitive}), tnode
                ]
            ]
            [Polarization and Disagreement\\
            (Section~\ref{sec:pd}), obj,
            child anchor=west,
            parent anchor=east,
                [Polarization\\
                (Section~\ref{subsec:pd_polarization}), obj_det 
                ]
                [Disagreement\\
                (Section~\ref{subsec:pd_disagreement}), obj_det
                ]
                [Polarization and Disagreement\\
                (Section~\ref{subsec:pd_pd}),obj_det 
                ]
            ]
            [Other Objectives\\
            (Section~\ref{sec:others}), obj,
            child anchor=west,
            parent anchor=east,
                [Objectives Related to Overall Opinions\\
                (Section~\ref{subsec:others_overall}), obj_det
                ]
                [Objectives Related to Polarization and Disagreement\\
                (Section~\ref{subsec:others_pd}), obj_det
                ]
                [Additional Objectives\\
                (Section~\ref{subsec:others_additional}), obj_det
                ]
            ]
            ]
        ]
    \end{forest}
    }
    \caption{A high-level map of the main part of this survey. Blue boxes denote the objectives being optimized, while red boxes represent intervention mechanisms.}
    \label{fig:map}
\end{figure}

\section{Opinion-Dynamics Models}\label{sec:models}

In this section, we review opinion-dynamics models that have been frequently used to develop algorithmic interventions.
Since the primary focus of this survey is not the opinion-dynamics models themselves, we keep this section as concise as possible. Throughout the survey, we use $n$ and $m$ to denote the numbers of nodes and edges of a graph, respectively. 

\subsection{DeGroot Model}\label{subsec:DeGroot}
The \emph{DeGroot model} (also known as the \emph{French--DeGroot model})~\cite{DeGroot74} is one of the most fundamental opinion-dynamics models in the literature.
Let $G=(V,E,w)$ be a directed graph with positive edge weights, where an edge $(i,j)\in E$ indicates that $i$ pays attention to $j$ (e.g., $i$ follows $j$ on $\mathbb{X}$) and is therefore influenced by $j$. 
An undirected graph can be handled naturally by regarding each undirected edge $\{i,j\}$ as two directed edges $(i,j)$ and $(j,i)$. 
The weight function $w\colon E\to \mathbb{R}_{>0}$ quantifies the strength of this relationship (e.g., the extent of exposure in the feed).
Let $W=(w_{ij})_{i,j\in V}$ denote the weighted adjacency matrix of $G$, where $w_{ij}=w((i,j))$ if $(i,j)\in E$ and $w_{ij}=0$ otherwise.
Define the (row-)random-walk matrix $P$ by
$P_{ij}=\frac{w_{ij}}{\sum_{k\in V} w_{ik}}$ for $i,j\in V$.
By construction, $P$ is row-stochastic, i.e., $\sum_{j\in V} P_{ij}=1$ for every $i\in V$.
We are given an initial opinion vector $\bm{x}^{(0)}\in \mathbb{R}^{V}$, where $x^{(0)}_i$ denotes the initial opinion of node $i\in V$.

The dynamics evolve in discrete time. At each iteration, each node updates its opinion by taking a weighted average (i.e., a convex combination) of the opinions of the nodes it pays attention to. Specifically, for $t=1,2,\dots$, the opinion vector $\bm{x}^{(t)}\in \mathbb{R}^{V}$ is updated from $\bm{x}^{(t-1)}$ according to
\begin{align*}
\bm{x}^{(t)} = P\,\bm{x}^{(t-1)}.
\end{align*}
If $P$ is primitive, i.e., there exists some $k\in \mathbb{Z}_{>0}$ such that $P^k>O$ (namely $P^k_{ij}>0$ for any $i,j\in V$), then the limit of the opinion vector $\bm{x}^{(\infty)}\coloneqq \lim_{t\rightarrow \infty}\bm{x}^{(t)}$ exists and equals $(\bm{\pi}^\top\bm{x}^{(0)})\bm{1}$, where $\bm{\pi}\in \mathbb{R}_{>0}^V$ is the unique vector that satisfies $\bm{\pi}^\top P=\bm{\pi}^\top$ and $\sum_{i\in V}\pi_i=1$. Note therefore that in the limit vector, every node has the same opinion $\bm{\pi}^\top \bm{x}^{(0)}$, which is determined solely by $P$ and $\bm{x}^{(0)}$. 

In the literature of algorithmic interventions, the generalization of the DeGroot model called the \emph{Leader--Follower (LF) DeGroot model} has often been employed. In this model, the node set $V$ is divided into two disjoint sets: a leader set $L$ and a follower set $F$. The opinions of nodes in $L$, represented by $\bm{x}_L\in \mathbb{R}^L$, remain fixed over time, while the (initial) opinions of nodes in $F$, represented by $\bm{x}_F^{(0)}$, evolve in the same way as above. 
Specifically, for $t=1,2,\dots$, the opinion vector $\bm{x}_F^{(t)}\in \mathbb{R}^F$ is updated from $\bm{x}_F^{(t-1)}$ according to 
\begin{align*}
\bm{x}_F^{(t)}=P_{FF}\bm{x}_F^{(t-1)}+P_{FL}\bm{x}_L, 
\end{align*}
where $P_{FF}$ and $P_{FL}$ are block submatrices of $P$ collecting the follower-follower and follower-leader elements, respectively. 
If the spectral radius of $P_{FF}$, i.e., the maximum absolute value among eigenvalues of $P_{FF}$, is strictly less than $1$, the limit of the opinion vector $\bm{x}_F^{(\infty)}\coloneqq \lim_{t\rightarrow \infty}\bm{x}_F^{(t)}$ exists and equals $(I-P_{FF})^{-1}P_{FL}\bm{x}_L$. 
Note that in this model, followers' opinions in the limit may differ from one to another. 

\subsection{Friedkin--Johnsen Model}

The \emph{Friedkin--Johnsen (FJ) model} \cite{Friedkin+99} generalizes the DeGroot model. In this extension, each node possesses an innate opinion, which remains fixed, and an expressed opinion, which is updated by combining its innate opinion with the expressed opinions of its neighbors, depending on its susceptibility to persuasion. 
Therefore, in the FJ model, we are additionally given an innate-opinion vector $\bm{s}\in \mathbb{R}^V$, where $s_i$ represents an innate opinion of node $i\in V$ and a susceptibility vector $\bm{\alpha}\in [0,1)^V$, where $\alpha_i$ quantifies susceptibility to persuasion of node $i$: the higher the value of $\alpha_i$ is, the more susceptible the node $i$ is. 

The dynamics again evolve in discrete time. Specifically, the (expressed-)opinion vector $\bm{z}^{(t)}\in \mathbb{R}^V$ at time $t=1,2,\dots$ is determined based on $\bm{z}^{(t-1)}$ as follows: 
\begin{align*}
 \bm{z}^{(t)}=(I-\Lambda)\bm{s}+\Lambda P\bm{z}^{(t-1)},\quad 
 \bm{z}^{(0)}=\bm{s}, 
\end{align*}
where $\Lambda=\text{Diag}(\bm{\alpha})$ is the diagonal matrix where $\Lambda_{ii}=\alpha_i$ for $i\in V$.  
The limit of the opinion vector is given by 
\begin{align*}
  \bm{z}^{(\infty)}
  \coloneqq \lim_{t\rightarrow \infty} z^{(t)} = (I - \Lambda P)^{-1} (I-\Lambda) \bm{s}, 
\end{align*}
which exists since $\Lambda_{ii}=\alpha_i < 1$ for every $i\in V$. 

In the algorithmic-interventions literature, especially for the undirected-graph case, the term FJ model is often used to refer to the special case in which each node $i$ has the degree-dependent susceptibility value $\alpha_i=\frac{\sum_{j\in V}w_{ij}}{1+\sum_{j\in V}w_{ij}}$ (e.g., see \cite{Bindel+11,Bindel+15,Gionis+13}). Then, by simple calculation, the limit of the opinion vector can be written as
\begin{align*}
\bm{z}^{(\infty)}=(I+L)^{-1}\bm{s}, 
\end{align*}
where $L$ is the (combinatorial) graph Laplacian of $G$, i.e., $L=D-W$, where $D$ is the diagonal degree matrix with $D_{ii}=\sum_{j\in V}w_{ij}$ for $i\in V$~\cite{Chung97}. 
To adhere to the original definition of the FJ model, we reserve the term FJ model for the general case and refer to this degree-dependent-$\alpha$ case as the \emph{FJ model with degree-dependent susceptibility}.

\smallskip
\noindent
\textbf{Fast Computation of $\bm{z}^{(\infty)}$.}
Note that computing $\bm{z}^{(\infty)}$ in the FJ model with degree-dependent susceptibility takes $O(n^3)$ time using naive inversion even in undirected graphs, since it involves the matrix inverse $(I+L)^{-1}$. This is too slow in practice, and therefore faster methods for approximately computing $\bm{z}^{(\infty)}$ have been developed  by exploiting useful properties of the matrix $I+L$.
\citet{Xu+21_fast} noticed that a $(1+\epsilon)$-approximation of $\bm{z}^{(\infty)}$ (in terms of the $\ell_2$-norm) can be computed in time nearly-linear in $m$ by exploiting the Laplacian solver framework of \citet{spielman2014nearly}. The key observation is that computing $\bm{z}^{(\infty)}$ is equivalent to solving the linear system $(I+L) \bm{z}^{(\infty)}=\bm{s}$, where $I+L$ is symmetric and diagonally dominant, which allows to apply results from the Laplacian solvers literature. \citet{Xu+21_fast} also demonstrated that practical implementations of Laplacian solvers lead to highly efficient algorithms. 
\citet{Neumann+24} developed sublinear-time algorithms, which only make certain queries to the graph 
without having access to the entire graph but can approximate entries of $\bm{z}^{(\infty)}$ with small additive error. This was achieved by exploiting a connection between the FJ dynamics and Personalized PageRank~\cite{Page+99} observed by \citet{friedkin2014two} and by building upon ideas from sublinear-time PageRank algorithms~\cite{andersen2006local}.
Another line of work exploited the fact that $(I+L)^{-1}$ is the so-called \emph{forest matrix}~\cite{chebotarev1997matrix}. 
By building upon the classic algorithm for sampling spanning trees by \citet{Wilson96}, 
on directed graphs, \citet{Sun+24_2} showed how one can approximate diagonal elements of the forest matrix while 
\citet{Sun+24_1} considered off-diagonal elements.

\smallskip
\noindent
\textbf{Generalizations.} 
The FJ model has been extensively studied in computer science, and numerous generalizations have been proposed in the literature, some of which will be used in later sections.
Examples include generalizations of the interaction structure (signed networks~\cite{Rahaman+20,Zhou+24,Xu+21}, hypergraphs~\cite{Xu+24}, and higher-order neighbors~\cite{Zhang+20,Zhang+24}),
agent-level modeling (cognitive bias~\cite{Tsang+14} and limited information of neighbors' opinions~\cite{Fotakis+18,Fotakis+23}),
and exogenous or platform effects (information propagation~\cite{Tu+22}, external media sources~\cite{Out+24}, recommender systems~\cite{Sprenger+24}, and user-level timeline algorithms~\cite{Zhou+24_2}),
as well as extensions to multiple interdependent topics~\cite{parsegov2016novel}.

\subsection{Other Models}\label{subsec:models_others}

In addition to the DeGroot and FJ models, several other opinion-dynamics models will be used in later sections. 
A representative example is the voter model~\cite{Clifford+73,Holley+75}, in which each agent holds a binary opinion and, at random update times, adopts the opinion of a randomly chosen neighbor. 
This model captures opinion spread driven purely by local imitation and serves as a canonical baseline for studying consensus formation and long-run coexistence.
Another example is the bounded confidence model~\cite{Hegselmann+02,Bernardo+24}, where agents only interact with (and update toward) neighbors whose opinions lie within a prescribed confidence threshold. 
This simple homophily mechanism naturally yields fragmentation into multiple opinion clusters and can be used as a stylized model of polarization.
For a broader overview of opinion-dynamics models, we refer the reader to the recent comprehensive surveys~\cite{Peralta+25,Starnini+25,Shirzadi+25,Zareer+25_2}.

\section{Optimization of Overall Opinion}\label{sec:overall}

The seminal work of \citet{Gionis+13} initiated the study of \emph{opinion maximization}, aiming to maximize the overall opinion (i.e., sum of opinions), by distinguishing it from the traditional problem of influence maximization~\cite{Kempe+03,Kempe+15}. Specifically, they formulated the opinion maximization problem, referred to as \textsc{Campaign}, based on the FJ model with degree-dependent susceptibility, and focused on interventions that fix the expressed opinions of a limited number of nodes throughout the dynamics. 

Let $G=(V,E,w)$ be an edge-weighted undirected graph with an innate-opinion vector $\bm{s}\in [0,1]^V$.
The objective of their problem is to maximize the sum of opinions in the limit of the FJ model with degree-dependent susceptibility; they consider interventions that fix the expressed opinions of a target set $T\subseteq V$ to the maximum value of $1$ throughout the dynamics.
For $T\subseteq V$, let $g(T)$ denote the objective function.
To characterize the limit, the authors introduced an auxiliary edge-weighted directed graph $G'=(V',E',w')$ constructed from $G$ as follows.
Let $V'=V\cup \overline{V}$, where $\overline{V}=\{\overline{i}\mid i\in V\}$ is a copy of $V$.
Define $E'=\{(i,j),(j,i)\mid \{i,j\}\in E\}\cup \{(i,\overline{i})\mid i\in V,\, \overline{i}\in \overline{V}\}$,
with weights $w'_{ij}=w'_{ji}=w_{ij}$ for $\{i,j\}\in E$ and $w'_{i\overline{i}}=1$ for all $i\in V$.
Let $P'$ be the (row-)random-walk matrix on $G'$ defined in the same manner as in Section~\ref{sec:models}. 
Let $U=V\setminus T$ and $B=\overline{V}\cup T$.
Define $\bm{f}\in [0,1]^B$ by setting $f_{\overline{i}}=s_i$ for $\overline{i}\in \overline{V}$ and $f_j=1$ for $j\in T$.
Then the limit of the expressed opinions on $U$ admits the representation
$\bm{z}^{(\infty)} \coloneqq (I-P'_{UU})^{-1}P'_{UB}\bm{f}$,
where $P'_{UU}$ and $P'_{UB}$ are the $U$-$U$ and $U$-$B$ block submatrices of $P'$, respectively.
Since the expressed opinions of nodes in $T$ are fixed to $1$, the objective function can be written as $g(T)=\sum_{i\in U}z^{(\infty)}_i+|T|$. We formally introduce their problem: 

\begin{problem}[\textsc{Campaign}~\cite{Gionis+13}]\label{prob:campaign}
Given an edge-weighted undirected graph $G=(V,E,w)$, an innate-opinion vector $\bm{s}\in [0,1]^V$, and a budget $b\in \mathbb{Z}_{>0}$, the goal is to find a target set $T\subseteq V$ with $|T|\leq b$ that maximizes the overall opinion $g(T)$ of the limit of the expressed opinions in the FJ model with degree-dependent susceptibility.
\end{problem}

The authors proved that \textsc{Campaign} is NP-hard by constructing a polynomial-time reduction from the vertex cover problem on regular graphs~\cite{Garey+02}. On the positive side, they showed that the objective function $g(T)$ is monotone and submodular.
Let $S$ be a finite set. A function $f\colon 2^S \rightarrow \mathbb{R}$ is said to be \emph{monotone} if $f(X)\leq f(Y)$ holds for any $X\subseteq Y\subseteq S$. A function $f$ is said to be \emph{submodular} if for any $X,Y\subseteq S$, it satisfies $f(X)+f(Y)\geq f(X\cup Y)+f(X\cap Y)$~\cite{Fujishige05}. An equivalent definition of submodularity is the diminishing marginal returns property: $f$ is submodular if $f(X\cup \{s\}) - f(X) \geq f(Y\cup \{s\}) - f(Y)$ holds for any $X\subseteq Y\subseteq S$ and $s\in S$~\cite{Fujishige05}.
Therefore, despite the NP-hardness, \textsc{Campaign} can be seen as a special case of submodular maximization subject to a cardinality constraint, for which the standard greedy algorithm offers a $(1-1/\mathrm{e})$-approximation~\cite{Nemhauser+78}.

Specifically, the greedy algorithm for \textsc{Campaign} works as follows: it starts with an empty set $T=\emptyset$. At each iteration, it evaluates the objective value $g(T\cup \{v\})$ for every $v\in V\setminus T$, and selects the node $v^*$ that gives the highest increase. This process is repeated until the budget $b\in \mathbb{Z}_{>0}$ is exhausted. The time complexity of the algorithm is analyzed as follows: each iteration requires $O(n)$ evaluations of the objective function, each taking time $O(mI)$ when the power method is used for computing the limit, where $I$ is the number of iterations in the power method. Since the number of outer-iterations is $b$, the overall time complexity is $O(bmnI)$.
Although this algorithm achieves a $(1-1/\mathrm{e})$-approximation, it does not scale to large networks, even with the use of lazy evaluation techniques for submodular maximization~\cite{Minoux05}. 

To address this limitation, the authors proposed five scalable heuristics: \texttt{Degree}, \texttt{FreeDegree}, \texttt{RWR}, \texttt{Min-S}, and \texttt{Min-Z}. These methods are motivated by the empirical observation that the node selection order in the greedy algorithm is highly correlated with the descending order of weighted degrees and the ascending order of innate opinions.
\texttt{Degree} selects the top-$b$ nodes based on weighted degree, and runs in $O(m+n\log n)$ time.
\texttt{FreeDegree} is a greedy-style algorithm that, at each step, selects the node with the highest sum of edge weights to neighbors not yet selected. This algorithm runs in $O(m+bn)$ time. 
\texttt{RWR} (Random Walk with Restart) considers both weighted degrees and innate opinions. It performs a random walk on $G$ with a restart probability proportional to $\max_{v\in V}s_v - s_i$ for node $i$, selects the top node by stationary distribution, and repeats the process with appropriate updates. This method has a time complexity of $O(bmI)$ when using the power method.
\texttt{Min-S} selects the top-$b$ nodes with the smallest innate opinions and runs in $O(m+n\log n)$ time.
\texttt{Min-Z} iteratively selects the node with the smallest expressed opinion at the limit, running in $O(bmI)$ time.
Computational experiments show that the greedy algorithm performs best on small networks, while three of the heuristics, \texttt{Degree}, \texttt{FreeDegree}, and \texttt{RWR}, achieve comparable performance with much greater scalability.

The work of \citet{Gionis+13} has inspired a large body of follow-up research.
In the following, we review this subsequent literature by categorizing studies according to the types of intervention mechanisms they consider.
Table~\ref{tab:overall_summary} provides an overview of the literature.

\begin{table}[t]
\caption{Summary of the literature on optimizing overall opinion. In the Graph column, the first character (U/W/S) indicates edges are unweighted/weighted/signed, and the second character (U/D) indicates edges are undirected/directed. In the Opinion-dynamics model column, FJ (deg) and FJ (gen) represent the FJ model with degree-dependent susceptibility and the general FJ model, respectively. If a paper considers multiple intervention mechanisms, we categorize it under the mechanism it primarily focuses on.}
\label{tab:overall_summary}
\resizebox{\textwidth}{!}{
\begin{tabular}{ccccccc}
\toprule
Intervention & Reference & Graph & Opinion-dynamics model & Main results \\
\midrule
\makecell[c]{Expressed opinions\\ (Section~\ref{sec:overall} intro.)} & \citet{Gionis+13} &W--U& FJ (deg) &Problem~\ref{prob:campaign} \& Approx./Heuristic alg. \\
\midrule
Innate opinions & \citet{Ahmadinejad+14,Ahmadinejad+15} &W--D& FJ (deg) &Problems~\ref{prob:overall_influencing}--\ref{prob:budgeted_stochastic_influencing} \& Exact/Approx. alg. \\
(Section~\ref{subsec:overall_innate})   & \citet{Sun+23} & U--D & FJ (deg) & Approx. alg.& \\
    & \citet{Wang+25} & U--D & FJ (gen) & Exact/Approx. alg.\\
    & \citet{Xu+21} &S--D  & FJ (deg) on signed networks & Exact alg.\\
    & \citet{He+22} &S--D  & FJ (gen) on signed networks & Exact alg.\\
    & \citet{Li+22} &W--D & FJ (gen) &Exact alg. \\
\midrule
Leader selection & \citet{Yildiz+13}    &U--D & Binary voter & Approx. alg.\\
(Section~\ref{subsec:overall_leader})   & \citet{Hunter+22} &W--D & LF DeGroot & Problem~\ref{prob:stubborn_agent_placement} \& Approx. alg.\\
    & \citet{Yi+21} &W--D & LF DeGroot & Problem~\ref{prob:absolute_leader_selection} \& Approx. alg. \\
    & \citet{Zhou+23_3} &U--U & LF DeGroot & Approx. alg. (acceleration) \\
    & \citet{Zhu+24} &W--U & DeGroot with external leaders & Approx. alg.\\
    & \citet{Hudson+21} &U--D &BIC & Heuristic alg. \\
    & \citet{Wan+25} &U--D &BIC & Heuristic alg. \\
\midrule
Network structure & \citet{Zhou+21} &U--U &LF DeGroot &Problem~\ref{prob:opinion_maximization_edge_addition} \& Approx. alg. \\
(Section~\ref{subsec:overall_structure})  & \citet{Zhou+23_1} & U--D & LF DeGroot & Approx. alg.\\
    & \citet{Zhu+25} &U--U & LF FJ (deg) & Approx. alg. \\
    & \citet{Gong+20} &U--D & FJ (gen) with bounded confidence & Heuristic alg.\\
\midrule
Susceptibility to persuasion & \citet{Abebe+18} & U--U & FJ (gen) & Problem~\ref{prob:opinion_susceptibility} \& Heuristic alg.\\
(Section~\ref{subsec:overall_susceptibility})  & \citet{Chan+19} &W--U & FJ (gen) & Exact alg.\\
    & \citet{Abebe+21} &W--U & FJ (gen) & Exact/Heuristic alg. \\
    & \citet{Marumo+21} &W--U & FJ (gen) & Problem~\ref{prob:opinion_susceptibility_with_constraint} \& Heuristic alg.\\
    & \citet{Chan+21} &W--U & FJ (gen) & Problem~\ref{prob:opinion_susceptibility_with_constraint} ($p=1$) \& NP-hardness \\
\midrule
Multi-stage & \citet{Liu+18} & W--D & Multi-round LT with ratings &  Heuristic alg.\\
(Section~\ref{subsec:overall_multi})  & \citet{Alla+23} & W--D & DeGroot on signed networks & Heuristic alg.\\
    & \citet{He+21} & S--D& DeGroot on signed networks & Heuristic alg. \\
        & \citet{He+21_2,He+20} & W--U& Weighted voter model &  Heuristic alg. \\
    & \citet{He+23} & W--D & Adaptive cooperation model based on Q-learning &Heuristic alg. \\
    & \citet{Nayak+19} &U--U & Dirichlet Bayesian Network&  Heuristic alg.\\
    & \citet{Zhang+25} & W--D & LF DeGroot & Approx./Heuristic  alg.  \\
    & \citet{Wang+25} &S--D &FJ (deg) with silent nodes &Heuristic alg. \\
\midrule 
Competitive environments & \citet{grabisch2018strategic} & W--D & LF DeGroot & Nash equilibrium analysis\\
(Section~\ref{subsec:overall_competitive})  & \citet{Dhamal+19,Dhamal+20} & W--D& FJ (gen) with direct investment/campaign weights &  Stackelberg/Nash equilibrium analysis\\
    & \citet{Chen+24} & W--D& FJ (gen) & No-regret strategy for approx. minimax equilibrium\\
    & \citet{Bastopcu+25} & W--U & LF DeGroot & Nash equilibrium analysis\\
    & \citet{He+22_2} & S--D & Activated dynamic opinion model with Q-learning &Heuristic alg. \\
\bottomrule
\end{tabular}
}
\end{table}

\subsection{Interventions on Innate Opinions}\label{subsec:overall_innate}

\citet{Ahmadinejad+14,Ahmadinejad+15} formulated a variety of problems of opinion maximization in the FJ model with degree-dependent susceptibility, with a particular focus on the interventions that modify the innate opinions of a limited number of nodes. 
For an innate-opinion vector $\bm{s}\in [0,1]^V$, let $\bm{z}^{(\infty)}(\bm{s})$ denote the limit of the opinion vector of the FJ model with degree-dependent susceptibility, i.e., $\bm{z}^{(\infty)}(\bm{s})=(I+L)^{-1}\bm{s}$. 
The authors introduced four optimization problems, each of which has two different types of constraints (or objectives). 

\begin{problem}[\textsc{Overall Influencing}~\cite{Ahmadinejad+14,Ahmadinejad+15}]\label{prob:overall_influencing}
Given an edge-weighted directed graph $G=(V,E,w)$, an initial innate-opinion vector $\bm{s}\in [0,1]^V$, and a budget $b\in \mathbb{R}_{>0}$, the goal is to find a modified innate-opinion vector $\bm{s}'\in [0,1]^V$ that maximizes the overall opinion $g(\bm{s}')\coloneqq \sum_{v\in V}z^{(\infty)}(\bm{s}')_v$ subject to either the fractional budget constraint $\|\bm{s}-\bm{s}'\|_1\leq b$ or the integral budget constraint $\|\bm{s}-\bm{s}'\|_0\leq b$. 
\end{problem}

The authors first studied \textsc{Overall Influencing} and designed a polynomial-time exact algorithm for both the fractional and integral budget constraint cases. 
To this end, they showed that the unit increase in the innate opinion of a node $v$ results in the increase in the overall opinion by the column sum corresponding to $v$ in the matrix $(I+L)^{-1}$. 
The algorithm iteratively increases the innate opinion of a node that results in the maximum increase in the overall opinion until it reaches $1$ (or the maximum value within the budget when we handle the fractional budget constraint). The time complexity of the algorithm is dominated by computing the matrix inverse $(I+L)^{-1}$, which can be done in $O(n^\omega)$ time in theory, where $\omega$ is the matrix multiplication exponent, or in $O(n^3)$ time using naive inversion. 

\begin{problem}[\textsc{Targeted Influencing}~\cite{Ahmadinejad+14,Ahmadinejad+15}]\label{prob:targeted_influencing}
Given an edge-weighted directed graph $G=(V,E,w)$, an initial innate-opinion vector $\bm{s}\in [0,1]^V$, 
a set of target nodes $T\subseteq V$, and a threshold vector $\bm{t}\in \mathbb{Z}_{\geq 0}^T$, 
the goal is to find a modified innate-opinion vector $\bm{s}'\in [0,1]^V$ that minimizes either the fractional cost $\|\bm{s}-\bm{s}'\|_1$ or the integral cost $\|\bm{s}-\bm{s}'\|_0$
subject to the constraint $\bm{z}^{(\infty)}(\bm{s}')\geq \bm{t}$. 
\end{problem}

For \textsc{Targeted Influencing} with the fractional cost, the authors gave a linear programming formulation, which computes an optimal solution in polynomial time. 
For \textsc{Targeted Influencing} with the integral cost, they proved that the problem is hard to approximate within a factor of $o(\log n)$ unless $\mathrm{P}=\mathrm{NP}$ by constructing a polynomial-time approximation-preserving reduction from the set cover problem~\cite{Garey+02}. On the positive side, for any $\epsilon >0$, they designed a polynomial-time $O(\log (n/\epsilon))$-approximation algorithm that might violate the constraint but guarantees $z^{(\infty)}(\bm{s}')_v\geq t_v - \epsilon n$ for every $v\in T$. 

\begin{problem}[\textsc{Budgeted Influencing}~\cite{Ahmadinejad+14,Ahmadinejad+15}]\label{prob:budgeted_influencing}
Given an edge-weighted directed graph $G=(V,E,w)$, an initial innate-opinion vector $\bm{s}\in [0,1]^V$, a threshold vector $\bm{t}\in \mathbb{Z}_{\geq 0}^V$, and a budget $b\in \mathbb{Z}_{>0}$, the goal is to find a modified innate-opinion vector $\bm{s}'\in [0,1]^V$ that maximizes the number of nodes $v\in V$ that satisfy $z^{(\infty)}(\bm{s}')_v\geq t_v$ subject to either the fractional or integral budget constraint. 
\end{problem}

 The authors proved that \textsc{Budgeted Influencing} is at least as hard to approximate as the Densest $k$-Subgraph problem (D$k$S)~\cite{Lanciano+24} up to a constant factor, 
meaning that we cannot expect any algorithm with a practical approximation ratio. 
Indeed, under some computational complexity assumption, D$k$S cannot be approximated within a factor of $n^{\frac{1}{(\log \log n)^c}}$ for some $c >0$, and the current best approximation ratio is $O(n^{1/4+\epsilon})$ for any $\epsilon>0$~\cite{Lanciano+24}. 

\begin{problem}[\textsc{Budgeted Stochastic Influencing}~\cite{Ahmadinejad+14,Ahmadinejad+15}]\label{prob:budgeted_stochastic_influencing}
We are given an edge-weighted directed graph $G=(V,E,w)$ with an initial innate-opinion vector $\bm{s}\in [0,1]^V$. Assume that each $v\in V$ has a threshold $t_v$ that is not fixed but uniformly distributed in $[0,1]$. The goal is to find a modified innate-opinion vector $\bm{s}'\in [0,1]^V$ that maximizes the expected number of nodes $v\in V$ that satisfy $z^{(\infty)}(\bm{s}')_v\geq t_v$ subject to either the fractional or integral budget constraint. 
\end{problem}

The authors finally investigated \textsc{Budgeted Stochastic Influencing}, a stochastic variant of \textsc{Budgeted Influencing}, and proved that owing to the expectation in the objective, the problem is reducible to \textsc{Overall Influencing} and thus polynomial-time solvable using the greedy algorithm. 

Later, \citet{Sun+23} studied the minimization counterpart of \textsc{Overall Influencing} with the integral budget constraint on unweighted directed graphs $G=(V,E)$. 
Since this problem is equivalent to \textsc{Overall Influencing}, it is solvable in $O(n^\omega)$ time~\cite{Ahmadinejad+14,Ahmadinejad+15}. 
The main contribution by \citet{Sun+23} is to design a more scalable algorithm. 
To this end, they interpreted the overall opinion from the perspective of spanning converging forests~\cite{Agaev+01}. 
Based on this and Wilson’s algorithm~\cite{Wilson96}, they proposed a spanning-forest-sampling-based algorithm. 
The algorithm has an additive approximation guarantee and has a time complexity of $O(m+\ell n)$, where $\ell$ is the number of samples. 
Finally, the authors performed extensive experiments on real-world networks, 
demonstrating that their algorithm outputs near-optimal solutions much faster than the naive exact algorithm. 

Recently, \citet{Wang+25} considered the problem called \textsc{OpinionMax}, a generalization of \textsc{Overall Influencing} with the integral budget constraint on unweighted directed graphs $G=(V,E)$, where the general FJ model is employed.  
Recall that the limit of the opinion vector is written as $\bm{z}^{(\infty)} = (I - \Lambda P)^{-1} (I-\Lambda) \bm{s}$. 
Then, it is easy to see that the problem is still solvable by repeatedly selecting the node with the largest column sum in the matrix 
$(I - \Lambda P)^{-1} (I-\Lambda)$ until the budget is exhausted, resulting in a time complexity of $O(n^\omega)$. 
To reduce this, the authors designed two efficient sampling-based algorithms: one based on random walks and the other based on forest sampling. 
The time complexity of these methods is analyzed theoretically, and their efficiency is also demonstrated experimentally.

\citet{Xu+21} introduced a generalization of \textsc{Overall Influencing} with the fractional budget constraint, where the FJ model with degree-dependent susceptibility on signed networks is employed. 
The authors showed that the problem is still polynomial-time solvable using the above greedy-type algorithm. 
They also introduced an expressed-opinion counterpart of the problem, proved its NP-hardness, and proposed a greedy algorithm. 
Later, \citet{He+22} further generalized these problems to the ones with the general FJ model on signed networks, and proposed a scalable ADMM (Alternating Direction Method of Multipliers) algorithm.
\citet{Li+22} introduced an opinion minimization problem under the general FJ model, in which the network structure may be subject to perturbations by some malicious users.

\subsection{Interventions on Leader Selection}\label{subsec:overall_leader}

Several studies have investigated opinion maximization and minimization with the intervention of locating leaders or stubborn agents. \citet{Yildiz+13} initiated the study of this type of problem: they considered the binary voter model~\cite{Holley+75} with stubborn agents. In this model, some agents are stubborn and never change their opinions from either $0$ or $1$, while the other agents stochastically update their opinions in $\{0,1\}$ based on their neighbors' opinions. The authors introduced an opinion maximization problem under this model, where the intervention consists of selecting a limited number of stubborn agents with opinion $1$, and proposed the greedy $(1-1/\mathrm{e})$-approximation algorithm. 

\citet{Hunter+22} studied opinion maximization under the LF DeGroot model with interventions that select leaders (which they also call stubborn agents). 
For a leader set $L\subseteq V$, let $\bm{x}_F^{(\infty)}(L)$ be the limit of the opinion vector on the followers $F=V\setminus L$ under the LF DeGroot model. Then their problem is formulated as follows: 
\begin{problem}[Stubborn Agent Placement~\cite{Hunter+22}]\label{prob:stubborn_agent_placement}
Given an edge-weighted directed graph $G=(V,E,w)$, an initial opinion vector $\bm{x}^{(0)}\in [0,1]^V$, and a budget $b\in \mathbb{Z}_{>0}$, the goal is to find a leader set $L\subseteq V$ with opinion $1$ with $|L|\leq b$ that maximizes the followers' overall opinion $\sum_{v\in V\setminus L} x_F^{(\infty)}(L)_v$ of the LF DeGroot model. 
\end{problem}

The authors presented the greedy $(1 - 1/\mathrm{e})$-approximation algorithm. To improve the scalability of this algorithm, they also proposed a method for identifying potential candidate elements to be considered in each iteration of the algorithm, based on the concept called harmonic influence centrality.

\citet{Yi+21} then investigated opinion maximization and minimization under the LF DeGroot model with two groups of leaders: 
\begin{problem}[Absolute Leader Selection~\cite{Yi+21}]\label{prob:absolute_leader_selection}
Given a strongly-connected edge-weighted directed graph $G=(V,E,w)$, an initial opinion vector $\bm{x}^{(0)}\in [0,1]^V$ with a nonempty set $L_0\subseteq V$ of $0$-leaders (with opinions $0$), a candidate set $Q\subseteq V\setminus L_0$, a target value $t \in [0,1]$, and a budget $b\in \mathbb{Z}_{>0}$, the goal is to find a set $L_1\subseteq Q$ of $1$-leaders (with opinions $1$) with $|L_1|\leq b$ that minimizes the absolute difference between the average opinion (over all nodes) in the limit of the LF DeGroot model and the target value $t$. 
\end{problem}

The authors first showed that the problem is NP-hard.
The special case of $t = 1$ can be seen as an opinion maximization problem. They then proved that the average opinion at the limit is monotone and submodular with respect to $L_1$. Based on this observation, they mentioned the greedy $(1 - 1/\mathrm{e})$-approximation algorithm for maximizing the average opinion.
For the general case of $t \in [0,1]$, the authors reduced the problem to a particular submodular optimization problem and proposed a sophisticated greedy algorithm.
The authors also studied a problem under a variant of the LF DeGroot model, which they call the DeGroot model with influenced leaders, 
and presented analogous results for this problem. 

Later, \citet{Zhou+23_3} addressed the scalability issue of the naive greedy algorithm mentioned by \citet{Yi+21}, focusing on the case of $t = 1$. The authors expressed the average opinion at the limit using the pseudoinverse and the Schur complement of the graph Laplacian. They then approximated the term involving the pseudoinverse by Laplacian solvers and evaluated the Schur complement using node sparsifiers, which interpret the Schur complement as random walks with a specific average length~$\ell$. The time complexity of their algorithm is $O(\epsilon^{-2}m b \ell \log n)$, where $\epsilon \in (0,1/2]$ is an additive error. Their experiments demonstrate that the algorithm obtains a solution comparable to \citet{Yi+21} more efficiently.
\citet{Zhu+24} studied a variant of Absolute Leader Selection, where there are two external leaders (an adversarial leader and a defending leader with opposite purposes) and one is asked to select a limited number of direct followers of the defending leader that minimizes the influence of the adversarial leader. 
\citet{Hudson+21} introduced the novel opinion-dynamics model, Behavioral Independent Cascade (BIC) model, which incorporates individuals’ personalities, and studied the leader-selection problem under this model.
\citet{Wan+25} later addressed the problem of finding multiple leader sets (rather than a single set) under the BIC model, for robust decision-making.

\subsection{Interventions on Network Structure}\label{subsec:overall_structure}

\citet{Zhou+21} formulated an opinion maximization problem based on the LF DeGroot model, with a particular focus on the interventions on network structure. 
For an unweighted undirected graph $G=(V,E)$, let $\bm{x}_F^{(\infty)}(G)$ be the limit of the opinion vector on the followers $F\subseteq V$ under the LF DeGroot model. 
The authors introduced the following problem (and its minimization counterpart): 
\begin{problem}[Opinion Maximization~\cite{Zhou+21}]\label{prob:opinion_maximization_edge_addition}
Given an unweighted undirected connected graph $G=(V,E)$ with a $0$-leader set $L_0\subseteq V$ and a $1$-leader set $L_1\subseteq V$, a candidate set of edges $Q\nsubseteq E$ between the $1$-leaders and the followers, and a budget $b\in \mathbb{Z}_{>0}$, the goal is to find $R\subseteq Q$ with $|R|\leq b$ such that $G_R=(V,E\cup R)$ has the maximum increase in the followers' overall opinion, i.e., $\sum_{v\in V\setminus (L_0\cup L_1)}x_F^{(\infty)}(G_R)-\sum_{v\in V\setminus (L_0\cup L_1)}x_F^{(\infty)}(G)$. 
\end{problem}

The authors first proved that the objective function of the above problem is monotone and submodular, demonstrating that the standard greedy algorithm admits a $(1-1/\mathrm{e})$-approximation. 
However, the naive implementation of the greedy algorithm results in the time complexity of $O(b|Q|n^3)$, which is only applicable to small networks. The authors showed that by seeing a single edge addition as a rank-$1$ update to the inverse matrix in the objective function, the time complexity in each iteration can be reduced from $O(|Q|n^3)$ to $O(|Q|n^2)$, resulting in the overall complexity of $O(b|Q|n^2)$, which is still prohibitive for large-scale networks. To overcome this issue, the authors further provided efficient approximate computation of the objective function, based on the property called the Symmetric, Diagonally-Dominant M-matrix, and developed $(1-1/\mathrm{e}-\epsilon)$-approximation algorithm that runs in $\widetilde{O}(km\epsilon^{-2})$ time\footnote{The $\widetilde{O}(\cdot)$ notation suppresses polylogarithmic factors.}, for any $\epsilon \in (0,1/2)$. 
Computational experiments demonstrate that their algorithm is accurate and scales to large networks. 

Later, \citet{Zhou+23_1} generalized the above problem to the setting of directed graphs. 
Based on the connection between the limit of opinions and absorbing probability of $\ell$-truncated absorbing random walks (similarly to \citet{Zhou+23_3}), 
the authors proposed a sampling-based $(1-1/\mathrm{e}-\epsilon)$-approximation algorithm running in 
$O(m+\epsilon^{-2}b|Q|\ell n\log n)$ time, where $\ell$ is the hyperparameter specifying the maximum length of random walks to be performed. 
The authors further accelerated the algorithm and achieved the time complexity of $O(\epsilon^{-3}b\ell n^{1/2}\log^{3/2} n)$ (except for the input time) using sample-materialization techniques. 
Extensive experiments demonstrate that 
the proposed algorithms achieve a solution quality comparable to the naive greedy algorithm while being significantly more efficient.

Recently, \citet{Zhu+25} extended Problem~\ref{prob:opinion_maximization_edge_addition} by replacing the underlying opinion-dynamics model from the DeGroot model to the FJ model, and derived analogous results to \citet{Zhou+21}. 
\citet{Gong+20} combined the FJ model with the Hegselmann--Krause (HK) model~\cite{Hegselmann+02} and, by focusing on structural hole spanners (i.e., nodes that bridge different communities), designed effective structural interventions to steer the overall opinion.

\subsection{Interventions on Susceptibility to Persuasion}\label{subsec:overall_susceptibility}

\citet{Abebe+18} formalized the problems of opinion minimization and maximization with the interventions at the level of susceptibility (rather than expressed or innate opinions). 
Consider the general FJ model. 
Let $\bm{\alpha}\in [0,1)^V$ be the susceptibility vector, where $\alpha_v$ quantifies susceptibility to persuasion of node $v$, 
and $g(\alpha)$ the overall opinion in the limit of the expressed opinions as a function of the susceptibility vector, i.e., $g(\bm{\alpha})
  \coloneqq \bm{1}^\top \bm z^{(\infty)}
  = \bm{1}^\top 
  (I - \Lambda P)^{-1} (I-\Lambda) \bm{s}$, where $\Lambda=\text{Diag}(\bm{\alpha})$ is the diagonal matrix where $\Lambda_{ii}=\alpha_i$ for $i\in V$. 
\citet{Abebe+18} formulated the following problem:
\begin{problem}[Opinion Susceptibility~\cite{Abebe+18}]\label{prob:opinion_susceptibility}
Given an unweighted undirected graph $G=(V,E)$, an innate-opinion vector $\bm{s}\in [0,1]^V$, and lower and upper bounds $\ell,u\in [0,1]$ on susceptibility values, 
the goal is to find a susceptibility vector $\bm{\alpha}\in [0,1)^V$ with $\ell \leq \alpha_v \leq u$ for all $v\in V$ that minimizes the overall opinion $g(\bm{\alpha})$ of the FJ model. 
The maximization counterpart can be defined in a similar manner. 
\end{problem}
Using toy examples, the authors showed that the interventions at the level of susceptibility may have much larger impact than those at the level of innate opinions. 
While the above problem is a continuous optimization problem, the authors revealed its combinatorial nature. 
Specifically, they proved that the problem has an optimal solution $\bm{\alpha}^*\in [0,1)^V$ such that $\alpha^*_v=\ell$ or $u$ for each $v\in V$. 
The proof is based on the random-walk interpretation of the FJ dynamics, conducted by \citet{Gionis+13}. 

In addition to the original problem, the authors also introduced the budgeted variant, where we have an initial susceptibility vector $\bm{\alpha}_\text{init}\in [0,1)^V$ and a budget $b\in \mathbb{Z}_{>0}$ and we can only change susceptibility values for at most $b$ nodes to optimize the overall opinion. The authors proved that this problem is NP-hard
and that the objective function is neither submodular nor supermodular. In light of the hardness result, they proposed a greedy algorithm, inspired by the aforementioned combinatorial nature. The algorithm starts with $T=\emptyset$ and iteratively adds $v$ to $T$ whose change of susceptibility value to $l$ or $u$ minimizes/maximizes the objective value. The algorithm can be implemented to run in $O(bn^3)$ time, using the Sherman--Morrison formula for computing the objective values. 

Later, \citet{Chan+19} revisited the (unbudgeted) problem. 
The authors focused only on the minimization problem, as the techniques required are essentially the same. 
They first generalized the previous problem to the case of edge-weighted undirected graphs together with possibly different lower and upper bounds $\ell_v, u_v$ on susceptibility values over nodes. 
They then thoroughly analyzed structural properties of the objective function. 
They first showed that the function (even on the unweighted graphs) is neither convex nor concave, contrary to the claim by \citet{Abebe+18} that the function is convex. 
On the other hand, they found the fact that any local optimum in terms of changing only a single susceptibility value is also a global optimum, in spite of its non-convexity. 
Using this fact, they designed a local search algorithm to find an optimal solution in polynomial time, which is also highly scalable in practice. 
The results of \citet{Abebe+18} and \citet{Chan+19} were later combined in \citet{Abebe+21}, which also provides a detailed discussion of heuristics for the budgeted problem.

\citet{Marumo+21} generalized the minimization problem to better handle real-world scenarios. 
In the original (unbudgeted) problems, we are allowed to determine the susceptibility vector $\bm{\alpha}\in [0,1)^V$ freely 
as long as it satisfies lower and upper bound constraints, i.e., $\bm{\alpha}\in [\bm{l}, \bm{u}]$ (i.e., $\alpha_v\in [\ell_v,u_v]$ for every $v\in V$). 
However, the obtained solution might not be implementable in real-world scenarios. 
Indeed, since the amount of changes of susceptibility is not considered, it would be too costly to change the susceptibility values for individuals based on the solution. 
To overcome this issue, \citet{Marumo+21} introduced the following optimization model that is able to limit the amount of changes of the susceptibility values in various forms: 
\begin{problem}[Opinion Susceptibility with $\ell_p$-Norm Constraint~\cite{Marumo+21}]\label{prob:opinion_susceptibility_with_constraint}
Let $p\in [1,\infty]$. Given an edge-weighted undirected graph $G=(V,E,w)$, an innate-opinion vector $\bm{s}\in [0,1]^V$, 
an initial susceptibility vector $\bm{\alpha}_\mathrm{init}\in [0,1)^V$, lower and upper bound vectors $\bm{l},\bm{u}\in [0,1]^V$ on susceptibility values, and a budget $b\in \mathbb{R}_{\geq 0}$, 
the goal is to find a susceptibility vector $\bm{\alpha}\in [0,1)^V$ that minimizes $g(\bm{\alpha})$ subject to the constraints that $\bm{\alpha}\in [\bm{l}, \bm{u}]$ and $\|\bm{\alpha}-\bm{\alpha}_\mathrm{init}\|_p\leq b$. 
\end{problem}
The budgeted problem of \citet{Abebe+18}, which just allows to modify susceptibility values for at most $b$ nodes, corresponds to the above problem with $p=0$.
Independent of \citet{Marumo+21}, Chan and Lee~\cite{Chan+21} introduced the above problem with $p=1$, for which they proved the NP-hardness. 
For the above problem, \citet{Marumo+21} designed an efficient projected gradient method that is applicable to million-scale graphs. 
It should be noted that since the above generalization has the $\ell_p$-norm constraint over all nodes, there is no longer the aforementioned combinatorial nature, 
which suggests using such a continuous optimization approach. 
The algorithm design is based on Nesterov's variant of the projected gradient methods~\cite{Nesterov13} and a reduction of the gradient computation to solving certain linear systems. 

\smallskip
\noindent
\textbf{Intervention-Related Insights.}
Motivated by the importance of characterizing the susceptibility to persuasion of users in several applications such as debiasing online surveys and finding social influencers, \citet{Das+14} addressed the problem of estimating the level of susceptibility to persuasion, using only the network structure and expressed opinions. They also addressed the \emph{seed recovery} problem, identifying the smallest subset of users that, when seeded initially with some non-neutral opinions, can accurately explain the current expressed opinions of users. 
They analyzed the problems and designed efficient algorithms for both of them, based on ideas from compressed sensing.

\subsection{Multi-Stage Interventions}\label{subsec:overall_multi}

While the interventions discussed so far primarily focus on optimizing the limit of opinions in the DeGroot or FJ models via one-shot actions, a growing body of work addresses multi-stage interventions.
\citet{Liu+18} proposed the Active Opinion Maximization (ATOM) problem as a sequential extension of \textsc{Campaign} (Problem~\ref{prob:campaign}) , which operates in a multi-round campaign where the users' opinions are initially unknown. They employed active learning to iteratively select seed nodes and refine opinion estimates over multiple rounds.
\citet{Alla+23} addressed opinion maximization in signed social networks, where edges represent trust (positive) or distrust (negative), using the Multi-Stage Linear Threshold (MSLT) model, inspired by the Linear Threshold (LT) model in influence maximization~\cite{Kempe+03,Kempe+15}. 
Their objective is to select a seed set $S\subseteq V$ that maximizes 
the overall opinion, using centrality measures and community structure.

He et al.~\cite{He+20,He+21,He+21_2,He+23} explored various multi-stage frameworks.
\citet{He+21} introduced an activated opinion maximization framework for signed networks, while \citet{He+20,He+21_2} proposed adaptive seeding strategies under a weighted voter model. 
\citet{He+23} later introduced the Dynamic Opinion Maximization Framework (DOMF), formulating the problem of maximizing rational opinions, which is the total positive opinions penalized by weighted negative ones, over a finite time horizon.
They modeled opinion formation using an adaptive cooperation model based on Q-learning theory. This allows nodes to dynamically learn and adjust their opinions (e.g., $+1, 0, -1$) based on rewards from neighbors.
The proposed seeding strategy is adaptively updated while aligning with the reinforcement learning process.

Other approaches tackle uncertainty using probabilistic models.
\citet{Nayak+19} formulated opinion maximization as a sequential decision problem using a Dynamic Bayesian Network (DBN). They represented opinions as time-varying Dirichlet distributions and proposed algorithms to optimize information spreading against adversarial sources over a finite horizon.
\citet{Zhang+25} investigated the impact of intervention timing under the LF DeGroot model. To maximize the overall opinion at a specific time via a limited number of interventions, they leveraged the objective's monotonicity and submodularity, and developed an efficient greedy algorithm and a time-importance-based heuristic.
\citet{Wang+2025LPMDP} studied an opinion maximization problem under an opinion-dynamics model that combines the FJ dynamics with silent nodes, and presented a multi-stage seed selection algorithm, which dynamically optimizes the seed distribution among communities.

\subsection{Interventions in Competitive Environments}\label{subsec:overall_competitive}

Several studies model interventions as competitive games between strategic agents.
\citet{grabisch2018strategic} analyzed a zero-sum game under the DeGroot model, where two strategic agents with fixed opinions compete to influence nonstrategic agents by establishing a single link, and characterized its Nash equilibrium.
\citet{Dhamal+19,Dhamal+20} defined competitive opinion optimization under an extension of the FJ model, where two camps allocate investments subject to budget constraints, providing Stackelberg and Nash equilibrium analyses.
Similarly, \citet{Chen+24} formulated a two-player zero-sum game under the FJ model where players directly modify individuals' innate opinions. They provided a no-regret learning strategy to achieve an approximate minimax equilibrium and analyzed multiagent learning convergence in the simultaneous game.

Recent work shifts to dynamic settings.
\citet{Bastopcu+25} studied a multi-round budget allocation problem where the network's initial state is stochastically drawn. They characterized closed-loop equilibrium strategies for multiple influencers competing under the DeGroot model, showing convergence to an approximate Nash equilibrium via online regret minimization.
\citet{He+22_2} considered signed social networks and aimed to maximize effective opinions under unknown opponent strategies.
Similarly to the approach in~\citet{He+23}, 
they proposed the activated dynamic opinion model based on stateless Q-learning.

\section{Optimization of Polarization and Disagreement}\label{sec:pd}

A large body of work has focused on mitigating harmful outcomes of opinion dynamics, particularly user polarization and conflict. In this section, we review the literature on optimizing---especially minimizing---objectives of polarization, disagreement, and their combinations. 
Table~\ref{tab:summary_pd} provides an overview of the literature. 

\begin{table}[t]
\caption{Summary of the literature on optimizing polarization and disagreement. 
Same notation and categorization rule as in Table~\ref{tab:overall_summary}.}
\label{tab:summary_pd}
\resizebox{\textwidth}{!}{
\begin{tabular}{cccccccc}
\toprule
Objective & Intervention & Reference & Graph & Opinion-dynamics model & Main results \\
\midrule
Polarization & Innate opinions & \citet{Matakos+17} &U--U & FJ (deg) & Problem~\ref{problem:matakos+17} \& Heuristic alg. \\
(Section~\ref{subsec:pd_polarization})  &                 & \citet{Rahaman+20} & S--D & FJ (gen) on signed networks & Exact alg. (continuous) \\
    & Network structure & \citet{Racz+23} &W--U & FJ (deg) & Heuristic alg. \\
    &                       & \citet{Xu+23_1} &W--U & Noisy LF DeGroot & Approx. alg. \\
    &                       & \citet{Kishore+24} &W--U   & FJ (deg) with two innate-opinion vectors & Heuristic alg. \\
    &                       & \citet{Wu+23} &U--U   & HK  & Heuristic alg. \\
    & Susceptibility to persuasion & \citet{Ristache+24} & W--U & FJ (gen) & Heuristic alg. \\
    & Competitive environments & \citet{Zhang+24_2} &W--U & FJ (deg) & Nash equilibrium analysis \\
\midrule
Disagreement & Network structure & \citet{Bindel+11,Bindel+15} &U--D & FJ (deg) & Problem~\ref{prob:pd_edge_addition} \& Exact/Approx. alg. \\
(Section~\ref{subsec:pd_disagreement}) & Leader selection  & \citet{Liu+25} &U--U & LF FJ (deg) & Approx. alg. \\
\midrule
Polarization & Innate opinions & \citet{Musco+18} &W--U & FJ (deg) & Problems~\ref{prob:pd_topo} and \ref{prob:pd_inn} \& Exact alg. \\
and          &    & \citet{Zhu+22} &U--U  & FJ (deg) & Approx. alg. (discrete) \\
Disagreement & Network structure & \citet{Zhu+21} &U--U & FJ (deg) & Problem~\ref{prob:pd_link_rec} \& Approx. alg. \\
(Section~\ref{subsec:pd_pd}) &  & \citet{Yi+20} &U--U & FJ (deg) & Approx. alg. \\
    & Rebalancing feeds & \citet{Cinus+23} &W--D & FJ (deg) & Problem~\ref{problem:Cinus+23} \& Heuristic alg. \\
    & Changing timelines  & \citet{Zhou+24_2} &W--U & Augmented FJ (deg) & Approx. alg. \\
    & Various types & \citet{Kuhne+25} &W--D & FJ (deg) & Heuristic alg. \\
    \cmidrule(lr){2-6}
(Limited Information) & Network structure & \citet{Chen+18} &W--U & FJ (deg) with unknown $\bm{s}$ & Problem~\ref{prob:pd_conflict} \& Heuristic alg. \\
    &    & \citet{Chaitanya+24} &W--U & FJ (deg) with unknown $\bm{s}$ & Heuristic alg. \\
    & Rebalancing feeds & \citet{Cinus+25} &W--D & FJ (deg) with unknown $\bm{s}$ & Problem~\ref{prob:pd_unknown} \& Heuristic alg. \\
    & Network structure & \citet{Cinus+25_2} &W--U & FJ (deg) with unknown $\bm{s}$ & Problem~\ref{prob:pd_bandit} \& Regret-guaranteed alg. \\
    \cmidrule(lr){2-6} 
(Maximization)    & Innate opinions & \citet{Gaitonde+20} &W--U & FJ (deg) & Problem~\ref{prob:pd_perturbations} \& Exact alg. \\
    &                 & \citet{Chen+22_2} &W--U & FJ (deg) & Heuristic alg. (discrete) \\
    &        & \citet{Tu+23} &W--U & FJ (deg) & Approx. alg. (discrete) \\
    &                & \citet{Zareer+25} &W--D & FJ (deg) & Heuristic alg. \\
    
\bottomrule
\end{tabular}
}
\end{table}

\subsection{Polarization}\label{subsec:pd_polarization}

The seminal work of \citet{Matakos+17} initiated the study of reducing polarization, by defining a novel polarization index for networks with opinion information. Let $G=(V,E,w)$ be an edge-weighted undirected graph. The authors considered the FJ model with degree-dependent susceptibility, where the innate-opinion vector is $\bm{s}\in [-1,1]^V$. An opinion value of $0$ indicates neutrality, and larger (resp. smaller) values indicate more positive (resp. more negative) opinions. They defined the polarization index for the limit $\bm{z}^{(\infty)}$ of the opinion vector of the FJ model as 
\begin{align*}
\pi(\bm{z}^{(\infty)})=\frac{\|\bm{z}^{(\infty)}\|_2}{n}, 
\end{align*}
which quantifies how far the opinions are from complete neutrality. 
Based on this index, the authors formulated a polarization minimization problem (on unweighted graphs) that seeks interventions on the innate opinions of a limited number of nodes. 
Recall that to emphasize the effect of the innate-opinion vector $\bm{s}$, we denote $\bm{z}^{(\infty)}$ by $\bm{z}^{(\infty)}(\bm{s})$. 
\begin{problem}[\textsc{ModerateInternal}~\cite{Matakos+17}]
\label{problem:matakos+17}
Given an unweighted undirected graph $G=(V,E)$, an innate-opinion vector $\bm{s}\in [0,1]^V$, and a budget $b\in \mathbb{Z}_{>0}$, the goal is to find a subset $T\subseteq V$ with $|T|\leq b$ such that the modified innate-opinion vector $\bm{s}'$ obtained by setting $s'_v=0$ for all $v\in T$ and $s'_v=s_v$ for all $v\in V\setminus T$ minimizes the polarization index $\pi(\bm{z}^{(\infty)}(\bm{s}'))$.
\end{problem}
The authors also introduced a counterpart of the problem, called \textsc{ModerateExpressed}, where the interventions are to fix expressed opinions of a limited number of nodes to $0$ throughout the dynamics, as in \textsc{Campaign} (Problem~\ref{prob:campaign}) of \citet{Gionis+13}. 
For both problems, they proved the NP-hardness, by constructing polynomial-time reductions from the $m$-subset sum problem and the vertex cover problem on regular graphs~\cite{Garey+02}, respectively. 

For \textsc{ModerateInternal}, the authors presented two algorithms: the Binary Orthogonal Matching Pursuit (\texttt{BOMP}) algorithm and the \texttt{GreedyInt} algorithm. 
The \texttt{BOMP} algorithm is inspired by the simple observation that \textsc{ModerateInternal} can be seen as a variant of sparse approximation~\cite{Natarajan95}, where one is asked to find a binary vector $\bm{x}\in \{0,1\}^V$ with $\|\bm{x}\|_0\leq b$ that minimizes $\|\bm{z}^{(\infty)}(\bm{s})-(I+L)^{-1}S\bm{x}\|_2^2$, where $L$ is the graph Laplacian of $G$ and $S=\text{Diag}(\bm{s})$. The algorithm starts with $\bm{x}=\bm{0}$ and iteratively changes the value from $0$ to $1$ for the element that minimizes the objective. The \texttt{GreedyInt} algorithm is the standard greedy algorithm, simply based on the original objective. Both algorithms run in $O(bn^2)$ time, assuming the pre-computation of the matrix $(I+L)^{-1}$. 
For \textsc{ModerateExpressed}, the authors presented the standard greedy algorithm, called \texttt{GreedyExt}. While the naive implementation is computationally expensive, utilizing the Sherman--Morrison formula reduces the time complexity to $O(bn^2)$. 
Computational experiments using real-world data, including a Twitter network from 2016 US Elections, demonstrate that the polarization index succeeds in capturing real-world opinion polarization and the proposed algorithms are effective in reducing it. Recently, \citet{Mylonas+25} introduced a fast method for computing the polarization index, using Graph Neural Networks~\cite{Scarselli+08}, and incorporated it into \texttt{GreedyExt} for acceleration. 

The work of \citet{Matakos+17} has inspired several follow-up studies.
\citet{Rahaman+20} introduced an extension of the FJ model that can handle (directed) signed networks and studied polarization minimization under this model via interventions on innate opinions with an $\ell_2$-norm constraint. Signed networks capture not only positive relationships and interactions but also antagonistic relationships and adversarial interactions. Their polarization minimization problem can be formulated as a second-order cone program~\cite{Alizadeh+03}. 

\citet{Racz+23} studied polarization minimization, focusing on interventions in network structure.
Their analysis starts from basic observations about the FJ dynamics, such as the fact that the dynamics contract polarization, and then connects these observations to structural graph properties. In particular, they showed that polarization is tightly controlled by the Cheeger constant~\cite{Chung97}: well-connected graphs (with a large Cheeger constant) exhibit small polarization relative to the innate opinions.
They then considered two optimization settings. In the full-information setting, where innate opinions are fixed, they formulated a nonconvex network design problem with an $\ell_0$-budget on edge-weight modifications. They analyzed this problem via a greedy, stepwise procedure, characterizing the effect of increasing a single edge weight at each step. This analysis shows that increasing an edge weight can either increase or decrease polarization, and it also provides sufficient conditions under which polarization is always reduced---for example, when two nodes have identical neighborhoods. These results motivate greedy heuristics based on coordinate descent and disagreement-seeking edge additions.
In the second, adversarial setting, innate opinions are chosen adversarially subject to a norm constraint. The resulting worst-case polarization admits a closed-form expression that depends on the Laplacian spectral gap, reducing the problem to maximizing this gap.

\citet{Xu+23_1} investigated polarization minimization under the noisy LF DeGroot model~\cite{Patterson+10}, in which leaders share a fixed common opinion and followers update their opinions following the standard DeGroot dynamics, but with additional white noise. The intervention consists of adding a limited number of edges incident to leaders. A key observation is that the polarization, defined as the steady-state variance of each follower’s deviation from the leaders’ opinion, equals half the sum of the effective resistance scores~\cite{Ghosh+08} between followers and the group of leaders. Building on this, the authors showed that the objective function is monotone and supermodular, and presented the standard greedy algorithm, along with an accelerated variant, to approximately maximize the reduction in the objective value.

\citet{Kishore+24} studied polarization minimization under the generalized FJ model where two innate-opinion vectors exist. Their motivation is compelling: although connecting nodes with opposing views is an obvious strategy for reducing polarization, such connections are often unrealistic and unlikely to lead to productive interactions. Hence, the authors introduced a two-dimensional model with two opinion vectors, each corresponding to a different topic. They proposed an intervention strategy that adjusts edge weights to more realistically reduce polarization on a primary issue by creating connections between individuals who share similarities on a separate, secondary topic.
\citet{Wu+23} investigated polarization minimization under the Hegselmann--Krause (HK)  model~\cite{Hegselmann+02}.
They formulated optimization problems in which polarization is quantified by summary statistics of the final opinion profile (e.g., the average absolute opinion), and examined interventions on both opinions and network structure.

\citet{Ristache+24} investigated interventions at the level of susceptibility, following the approach in \citet{Abebe+18,Abebe+21}. In their problem, given an edge-weighted undirected graph $G=(V,E,w)$, an innate-opinion vector $\bm{s}\in [0,1]^V$, a susceptibility vector $\bm{\alpha}\in [0,1]^V$, and a budget $b\in \mathbb{Z}_{>0}$, the goal is to find $T\subseteq V$ with $|T|\leq b$ such that the modified susceptibility vector $\bm{\alpha}'\in [0,1]^V$ obtained by setting $\alpha'_v=0$ or $1$ for every $v\in T$ and $\alpha'_v=\alpha_v$ for every $v\in V\setminus T$ either maximizes or minimizes the polarization (or a closely related quantity called the Mean Squared Error (MSE), which quantifies deviation from the average innate opinion). 
Here the authors assume the
existence of a unique equilibrium of the FJ dynamics. 
They showed that the above problem is NP-hard, regardless of the choice of objective, and that the objectives are neither submodular nor supermodular. They then presented an efficient heuristic based on fast recomputation of the limit of the expressed opinions together with lazy evaluations. Their experiments demonstrated that even with a small budget, the proposed heuristic can dramatically affect both polarization and MSE.

\citet{Zhang+24_2} analyzed a game defined under the FJ model with degree-dependent susceptibility, in which two players intervene in the innate opinions so as to maximize and minimize polarization, respectively.
The authors studied the functional properties of this game, characterized each player’s best response, and examined the properties of the resulting Nash equilibrium.

\smallskip
\noindent
\textbf{Intervention-Related Insights.}
\citet{Garimella+2017} studied polarization minimization without relying on opinion-dynamics models. Specifically, they proposed adding endorsement links between high-degree users across opposing partitions, and formulated this as selecting edges that most reduce the Random-Walk Controversy measure \cite{Garimella+18}.
The authors proposed an efficient and effective heuristic for the resulting NP-hard problem. 
\citet{Haddadan+21,Haddadan+22} addressed the risk that an individual becomes trapped in a polarized bubble by proposing structural interventions on the hyperlink graph that the individual navigates. They quantified this risk using the measure called Polarized Bubble Radius, 
and formulated risk reduction maximization as the problem of adding a limited number of inter-group edges. They showed that the objective is submodular and can be approximately optimized via an efficient greedy algorithm. 

Relatedly, \citet{fabbri2022rewiring} studied reducing radicalization pathways in recommendation graphs via edge rewiring. 
Similarly to \citet{Haddadan+21,Haddadan+22}, 
they modeled user navigation as a random walk on a directed out-regular graph whose nodes have binary, harmful/benign labels. They defined segregation as the expected time for a walk starting from harmful content to reach benign content under an absorbing-walk model, and showed that minimizing the maximum segregation with $b$ rewirings is NP-hard (and hard to approximate). To cope with this, they proposed a greedy heuristic that repeatedly solves the tractable $b=1$ case. 
Later, \citet{Coupette+23} generalized the model by allowing graded harmfulness via real-valued node costs and by dropping the assumption that benign nodes are absorbing. 
\citet{Adriaens+23} strengthened the theoretical foundations of this line of work. Specifically, they studied the direct minimization of hitting-time-based polarization measures via shortcut edge additions, rather than maximizing surrogate gain functions, and developed approximation algorithms.  
\citet{Interian+21} studied a related problem based on deterministic shortest-path distances and presented integer programming formulations. 
Recently, several studies considered reducing the visibility of specific nodes by deleting a limited number of edges, motivated by applications such as curbing the spread of misinformation and mitigating polarization \cite{Liu+24,Miyauchi+24}.

Another line of work aims to mitigate polarization through the design of article recommendation systems or influence campaigns that may encourage individuals in a social network to be aware of multiple, potentially diverse, viewpoints.
\citet{Aslay+18} and \citet{Matakos+22} considered maximizing diversity of exposure. They formulated the problem in the context of information propagation, as the task of recommending a small number of news articles to selected users.
\citet{Zareie+23} formulated the problem of maximizing the diversity of exposure by increasing susceptibility to persuasion of a limited number of nodes. 
\citet{Tu+20} studied the co-exposure maximization problem, which asks for two disjoint seed sets on a directed graph, each corresponding to a different campaign, so as to maximize the number of nodes activated by both campaigns.
Their problem is based on the Independent Cascade (IC) model in influence maximization~\cite{Kempe+03,Kempe+15} rather than an opinion-dynamics model. 
Later, \citet{Banaerjee+23} extended this problem by taking into account the competing nature of two campaigns, while \citet{Mao+24} modified the objective to maximize the number of edges whose endpoints are activated by multiple campaigns in total. 
\citet{Colin+24} introduced a generalized notion of diversity by incorporating the FJ dynamics, and studied a diversity maximization problem with interventions centered on a target node.
Note that, by contrast, \citet{Bail+2018} used Twitter data to show that exposure to opposing views on social media can increase political polarization.
\citet{Rawal+19} studied the problem of maximizing a notion of polarization via influence campaigns, aiming for raising awareness of controversial issues that are nonetheless under-discussed.

Complementing the above optimization-based approaches, several studies explored the structural drivers of polarization.
\citet{Santos+21} studied how link recommendation algorithms intervene in opinion dynamics and showed that recommending structurally similar users (i.e., users sharing many neighbors) increases clustering coefficient and makes networks more prone to polarization and radicalization, while recommending structurally dissimilar, cross-community links can mitigate polarization. 
Motivated by the fact that opinions in the real world are not governed deterministically by social influence,  \citet{Stern+21} introduced the noisy DeGroot and FJ models, which incorporate random opinion fluctuations, and examined various network configurations to ask what macroscopic-level characteristics (e.g., connectivity and community structures) promote or hinder polarization of opinions.
\citet{Chaturvedi+24} examined the question whether inter-group interactions around important events affect polarization between majority and minority groups in social networks, using the Global Reactions to COVID-19 on Twitter data. The authors found that for political and social events, inter-group interactions reduce polarization, while during communal events, inter-group interactions can increase group conformity.
\citet{Springsteen+24} investigated how several simple algorithmic filtering strategies for deciding whose opinions an individual is exposed to can impact polarization, using the opinion-dynamics model with skeptical agents~\cite{Tsang+14}. 
Empirical work by \citet{cinus2022effect}, and later by \citet{Morales+21}, examined how friend recommender systems affect polarization of opinions. 

By contrast, several studies investigated behavioral and cognitive drivers of polarization. 
Building on the geometric opinion-dynamics model~\cite{Hkazla+24}, \citet{Gaitonde+21} formalized properties such as sign-invariance, symmetry, and Markovian evolution, and characterized when strong or weak polarization emerges.
\citet{Chen+22} introduced a generalization of the FJ model in which each node is equipped with a confirmation-bias mechanism that can alter its listening structure. They analyzed the resulting dynamics and showed that, at equilibrium, the listening structure becomes polarized. 
\citet{Lim+22} demonstrated that opinion amplification causes extreme polarization in social networks, using the HK model~\cite{Hegselmann+02}. 
\citet{Rabb+23} performed simulations to see how parameters modeling audience fragmentation, selective exposure, and responsiveness of media agents to the beliefs of their subscribers influence polarization. They found that media ecosystem fragmentation and echo-chambers may not in themselves be as polarizing as initially postulated, in the absence of outside fixed media messages that are polarizing.
Recently, \citet{Ojer+25} introduced the multidimensional social compass model as a generalization of the DeGroot model, in which multiple topics can be considered and each node is endowed with a stubbornness parameter, potentially leading to polarization. They first examined the phase transition from polarization to consensus and identified a threshold value of social influence that governs this transition. They then proposed an edge-rewiring algorithm that can reduce the depolarization threshold.

\subsection{Disagreement}\label{subsec:pd_disagreement}

In addition to polarization, the objective called disagreement has also been studied. In general, for an edge-weighted undirected graph $G=(V,E,w)$ with opinion vector $\bm{z}\in \mathbb{R}^V$, the network disagreement index (NDI, or simply, disagreement) is defined as 
\begin{align*}
\eta(\bm{z})=\sum_{\{u,v\}\in E} w_{uv}(z_u-z_v)^2, 
\end{align*} 
quantifying the sum, over all edges, of the squared difference between endpoints' opinions, weighted by the edge weights. For directed graphs, the quantity is defined analogously by ignoring the direction (and multiplicity) of edges. 

The seminal work of \citet{Bindel+11,Bindel+15} initiated the study of optimization problems that minimize a quantity closely related to the disagreement.
They first interpreted the FJ model with degree-dependent susceptibility as a game in which each player (corresponding to a node $v$) incurs the cost
\begin{align*}
(z_v-s_v)^2 + \sum_{u\in N(v)}w_{uv}(z_u - z_v)^2,
\end{align*}
where $N(v)$ is the set of $v$'s neighbors, measuring the discrepancy between the current opinion $z_v$ and the innate opinion $s_v$ and neighbors’ opinions, and chooses a best response $z_v$ that minimizes this cost. 
Specifically, they showed the equivalence between the FJ model and this game, and in particular proved that the dynamics of this opinion game admit a unique Nash equilibrium, which coincides with the limit of the opinion vector under the FJ model, namely $\bm{z}^{(\infty)}=(I+L)^{-1}\bm{s}$. 
They then defined the social cost of an opinion vector $\bm{z}$ as the sum of the individual costs, i.e.,
\begin{align*}
c(\bm{z})\coloneqq \sum_{v\in V} \left((z_v-s_v)^2 + \sum_{u\in N(v)}w_{uv}(z_u - z_v)^2\right).
\end{align*}
They analyzed the price of anarchy defined as the ratio between the social cost at equilibrium and that under the optimal opinions minimizing the social cost. 
They proved that for the undirected case the price of anarchy is at most $9/8$, while for the directed case it can be unbounded (even on directed graphs with constant degrees). They then specified some classes of directed graphs, for which a constant price of anarchy is achievable. 
Later, \citet{Bhawalkar+13} analyzed the price of anarchy with respect to the generalized social cost, and \citet{Banihashem+25} more recently analyzed it under the generalized FJ model with multiple interdependent topics~\cite{parsegov2016novel}. 

\citet{Bindel+11,Bindel+15} also addressed minimizing the social cost at equilibrium through interventions in network structure.
Specifically, they studied the following problem:
\begin{problem}[Edge Addition for Social Cost Minimization~\cite{Bindel+11,Bindel+15}]\label{prob:pd_edge_addition}
Given an unweighted directed graph $G=(V,E)$ and an innate-opinion vector $\bm{s}\in [0,1]^V$, the goal is to find a set of edges whose addition minimizes the social cost at equilibrium under the FJ model, i.e., $c(\bm{z}^{(\infty)})$, 
subject to one of the following constraints: (i) we can add at most $b\in \mathbb{Z}_{>0}$ outgoing edges from a specific node $v\in V$; (ii) we can add at most $b\in \mathbb{Z}_{>0}$ incoming edges to a specific node $v\in V$; (iii) we can add at most $b\in \mathbb{Z}_{>0}$ edges over the graph; or (iv) no restrictions are imposed on the added edges. In case (i), the social cost is modified by excluding the individual cost associated with $v$. 
\end{problem}
They proved that cases (i), (ii), and (iii) are NP-hard via polynomial-time reductions from the subset sum problem, the minimum vertex cover problem, and D$k$S, respectively~\cite{Garey+02,Lanciano+24}. 
A positive result is a simple $9/4$-approximation algorithm for case (iv), which adds edges in the opposite direction whenever they do not already exist. 
They also studied a variant of the problem on edge-weighted graphs, in which an arbitrary weight can be assigned to a specific edge $(i,j)$ to minimize the social cost at equilibrium, and proved that this problem is polynomial-time solvable. 

Recently, \citet{Liu+25} studied disagreement minimization under the LF FJ model (i.e., an FJ counterpart of the LF DeGroot model), where leaders' innate and expressed opinions are fixed at $0$. They focused on leader placement as an intervention in a social network. Given an unweighted undirected connected graph $G=(V,E)$ and a budget $b\in \mathbb{Z}_{>0}$, the task is to choose a set of leaders $L\subseteq V$ with $|L|\leq b$ to minimize disagreement between leaders and followers. They showed that the objective is monotone and supermodular, and suggested using a sophisticated greedy algorithm~\cite{Qian+15} 
to obtain an approximate solution, along with a faster alternative.

\smallskip
\noindent
\textbf{Intervention-Related Insights.} 
\citet{Dandekar+13} showed that the DeGroot model does not amplify disagreement; that is, in the DeGroot model, disagreement is monotonically non-increasing throughout evolution.
The authors then generalized the DeGroot model by incorporating a well-known phenomenon in social psychology, called biased assimilation, which means that when individuals are presented with mixed or inconclusive evidence on a complex issue, they tend to draw undue support for their initial position, thereby arriving at a more extreme opinion.
They demonstrated that if individuals exhibit sufficiently strong bias, this generalized model amplifies disagreement.

\citet{Matakos+20} studied disagreement maximization, which they refer to as diversity maximization, aiming to break filter bubbles and echo chambers.
They did not consider opinion-dynamics models; instead, their setting can be viewed as intervening on opinions after the dynamics have evolved.
They formulated the task as a quadratic knapsack problem and proved inapproximability results.
They then proposed efficient algorithms inspired by existing approaches for quadratic knapsack, as well as scalable greedy heuristics.

\subsection{Polarization and Disagreement}\label{subsec:pd_pd}
The seminal work of \citet{Musco+18} highlighted the trade-off between polarization and disagreement under the FJ model with degree-dependent susceptibility: interventions designed to minimize polarization can inadvertently amplify disagreement, and vice versa.
Motivated by this observation, the authors initiated the study of polarization and disagreement minimization, aiming to minimize the two quantities simultaneously.
Let $G=(V,E,w)$ be an edge-weighted undirected graph and $\bm{s}\in [0,1]^V$ an innate-opinion vector. 
Let $\bm{z}^{(\infty)}$ be the limit of the opinion vector of the FJ model with degree-dependent susceptibility and  
$\overline{\bm{z}}^{(\infty)}$ the mean-centered vector of $\bm{z}^{(\infty)}$, i.e., $\overline{\bm{z}}^{(\infty)}=\bm{z}^{(\infty)}-\frac{\sum_{v\in V}z^{(\infty)}_v}{n}\bm{1}$.
They then defined the polarization as $\pi(\bm{z}^{(\infty)})=\|{\overline{\bm{z}}^{(\infty)}}\|_2^2$, which is equivalent to the polarization defined in \citet{Matakos+17}, except that here the vector is mean-centered to properly handle opinions in $[0,1]$ and the normalization by $n$ is removed.
The authors employed the aforementioned disagreement $\eta(\bm{z}^{(\infty)})$, which also equals the disagreement for the mean-centered vector $\overline{\bm{z}}^{(\infty)}$. The authors defined the polarization-disagreement index as the sum of the polarization and the disagreement, i.e., $\pi(\bm{z}^{(\infty)})+\eta(\bm{z}^{(\infty)})$, and formalized the following problems: 
\begin{problem}[Polarization and Disagreement Minimization over Graph Topologies~\cite{Musco+18}]\label{prob:pd_topo}
Given a set $V$ of nodes, an innate-opinion vector $\bm{s}\in [0,1]^V$, and a budget $b\in \mathbb{R}_{>0}$, 
the goal is to find an edge-weighted undirected connected graph on $V$ (or equivalently its graph Laplacian) 
that minimizes the polarization-disagreement index at the limit of the expressed opinions under the FJ model with degree-dependent susceptibility, i.e., 
$\pi(\bm{z}^{(\infty)}) + \eta(\bm{z}^{(\infty)})$,
subject to the constraint that the sum of edge weights is equal to $b$. 
\end{problem}
\begin{problem}[Polarization and Disagreement Minimization over Innate Opinions~\cite{Musco+18}]\label{prob:pd_inn}
Given an edge-weighted undirected graph $G=(V,E,w)$, an initial innate-opinion vector $\bm{s}\in [0,1]^V$, and a budget $b \in  \mathbb{R}_{>0}$, 
the goal is to find a modified innate-opinion vector $\bm{s}'\in [0,1]^V$ 
that minimizes the polarization-disagreement index at the limit of the expressed opinions under the FJ model with degree-dependent susceptibility, i.e., 
$\pi(\bm{z}^{(\infty)}) + \eta(\bm{z}^{(\infty)})$,
subject to the constraints that $\bm{s}'\leq \bm{s}$ and $\|\bm{s}'-\bm{s}\|_1 \leq b$. 
\end{problem}

As for the first problem, the authors did not intend to formulate an intervention problem; indeed, they are not given an input graph and allow to output any graph as long as it has the specified sum of edge weights. They are, therefore, more interested in purely computing the graph structure that minimizes the polarization-disagreement index. The authors first proved that the problem is convex programming problem by showing that both the objective function and the feasible region are convex, implying that the problem can be solved (within an arbitrary precision) in polynomial time~\cite{Boyd+04}. For the practical computational efficiency, they also derived a closed-form gradient. Finally, they proved that there always exists a graph with $O(n/\epsilon^2)$ edges that achieves a $(1+\epsilon +O(\epsilon^2))$-approximation, meaning that the polarization-disagreement index can be minimized by sparse graphs. 

The second problem is a genuine intervention problem. In this setting, only decreases to the original innate opinions are allowed, although the framework could accommodate other types of interventions (e.g., increasing innate opinions or, more generally, changing them within specified ranges). Similar to the above analysis, the authors showed that the problem can be formulated as a convex program and is therefore polynomial-time solvable.
Computational experiments demonstrate that the solutions to the above problems can significantly reduce the polarization-disagreement index in real-world scenarios, implying that real-world networks tend to have characteristics far away from minimizing polarization and disagreement. 

Later, \citet{Zhu+22} studied a variant of the second problem, which is equivalent to \textsc{ModerateInternal} (Problem~\ref{problem:matakos+17}) in \citet{Matakos+17} except that the objective can be the polarization-disagreement index. The authors showed that the objective is monotone and supermodular, and presented the greedy algorithm, achieving a $(1-1/\mathrm{e})$-approximation for the reduction in the objective. They then presented a faster greedy algorithm running in $\widetilde{O}(bm\epsilon^{-2})$ time that achieves a $(1-1/\mathrm{e}-\epsilon)$-approximation. Their speedup relied on efficiently approximating the objective value using the Johnson--Lindenstrauss Lemma~\cite{Johnson+84} and SDDM solvers, similarly to \citet{Zhou+21} and \citet{Zhu+21}. Their large-scale experiments demonstrated the effectiveness and efficiency of the proposed approach.

\citet{Zhu+21} studied polarization and disagreement minimization with interventions on network structure: 
\begin{problem}[Polarization and Disagreement Minimization via Link Recommendation~\cite{Zhu+21}]\label{prob:pd_link_rec}
Given an unweighted undirected connected graph $G=(V,E)$, an innate-opinion vector $\bm{s}\in [0,1]^V$, a candidate set of edges $Q\nsubseteq E$, and a budget $b\in \mathbb{Z}_{>0}$, 
the goal is to add at most $b$ edges in $Q$ to $G$ so as to minimize the polarization-disagreement index of the limit of the expressed opinions under the FJ model with degree-dependent susceptibility, i.e., $\pi(\bm{z}^{(\infty)})+\eta(\bm{z}^{(\infty)})$. 
\end{problem}
For this combinatorial problem, the authors first proved that the objective function is not submodular. 
However, despite the non-submodularity, they employed the standard greedy algorithm. 
The naive implementation causes the time complexity of $O(b|Q|n^3)$ while it can be reduced to $O(n^2(n+b|Q|))$ by seeing a single edge addition as a rank-$1$ update to the inverse matrix in the objective, as in \citet{Zhou+21}. 
It is known that the standard greedy algorithm for maximizing monotone non-submodular function with submodularity ratio $\gamma \in [0,1]$ and curvature $\alpha \in [0,1]$ achieves $\frac{1}{\alpha}(1-\mathrm{e}^{-\alpha \gamma})$-approximation~\cite{Bian+17}. 
By deriving the submodularity ratio and curvature of the objective, the authors derived an approximation ratio for the maximization counterpart of the problem, where the decrease in the polarization-disagreement index is maximized. 
They also designed a faster greedy algorithm running in $\widetilde{O}(bm\epsilon^{-2})$ time, by providing efficient approximate computation of the objective. 
\citet{Yi+20} investigated a similar problem under the LF DeGroot model, with interventions on network structure. 

\citet{Cinus+23} extended the work of \citet{Musco+18} to directed graphs. They studied polarization and disagreement minimization in the context of \emph{rebalancing social feed} on social media platforms. In their setting, an edge-weighted directed graph $G=(V,E,w)$ represents follower--followee relations, where a directed edge $(i,j)$ indicates that $i$ follows $j$ (i.e., $j$ can influence $i$). The edge weight $w_{ij}$ represents the strength of influence that $j$ exerts on $i$. They further assumed that the influence weights are normalized around every node so that each node has weighted out-degree $1$, i.e., $\sum_{v\in N_\text{out}(u)} w_{uv}=1$ for each $u\in V$, where $N_\text{out}(u)$ denotes the set of outgoing neighbors of $u$ (modeling the ``attention budget'' of a user).
The authors aimed to minimize polarization and disagreement by rebalancing the edge weights rather than modifying the network structure; that is, no new edges are introduced and the weighted out-degree of each node is preserved. Specifically, their problem is formulated as follows:
\begin{problem}[Polarization and Disagreement Minimization via Social-Feed Rebalancing~\cite{Cinus+23}]\label{problem:Cinus+23}
Given an edge-weighted directed graph $G=(V,E,w)$ with $\sum_{v\in N_\mathrm{out}(u)} w_{uv}=1$ for every $u\in V$ and an innate-opinion vector $\bm{s}\in [-1,1]^V$, the goal is to find a new edge weight $w'$ on $E$ that minimizes the polarization-disagreement index of the limit of the expressed opinions under the FJ model with degree-dependent susceptibility, subject to $\sum_{v\in N_\mathrm{out}(u)} w'_{uv}=1$ for every $u\in V$. 
\end{problem}
For this problem, the authors first proved that, when viewed as an optimization over (weighted adjacency) matrices, the objective is nonconvex while the feasible set is convex. They then designed a projected gradient method. They derived a closed-form expression for the gradient and showed how to compute it efficiently using the Biconjugate Gradient (BiCG) solver, similarly to \citet{Marumo+21}. They also showed that the projection step can be computed efficiently as a direct implication of the constraints. Finally, they formulated an undirected-graph counterpart, in which a doubly-stochastic weighted adjacency matrix is considered, and provided an algorithm based on semidefinite programming (SDP), in the same spirit of \citet{Musco+18}.

Relatedly, \citet{Zhou+24_2} studied how to reduce polarization and disagreement by making small changes to users’ timelines. Their model combines (i)~a fixed follower-followee graph with (ii)~an additional interaction layer induced by timeline exposure. To make this model computationally tractable, they assume that within the network there is only a small number of topics. Then, each user’s feed is described by a distribution over topics, and for each topic there is a distribution over which users’ content is shown. They showed that under this assumption, the additional interaction layer induces a low-rank modification of the follower adjacency structure, which means that the resulting opinion dynamics can still be simulated efficiently. They then showed how to optimize small, bounded adjustments to timeline compositions to minimize polarization and disagreement. 

Recently, \citet{Kuhne+25} developed a scalable hypergradient-based method for optimizing a range of interventions under the FJ model with degree-dependent susceptibility, including changes to the network structure and feed rebalancing. Their algorithm can accommodate arbitrary differentiable objective functions, such as polarization, disagreement, and the polarization-disagreement index.

The foundational work of \citet{Chen+18} initiated the study of polarization and disagreement minimization in a more realistic setting. Earlier intervention approaches based on the FJ model---not only for polarization and disagreement but also for other objectives---typically assume full access to nodes’ innate opinions. However, in practice, obtaining such information is inherently imprecise and costly. For example, inferring user beliefs on a controversial topic (e.g.,~COVID-19 vaccination or Brexit) on a social media platform would require either large-scale surveys or extensive analysis of user behavior (e.g., posts, reposts, and likes on platforms like $\mathbb{X}$). More fundamentally, even when such information is available for a given topic, one may wish to design interventions that are effective across multiple topics. 

The authors therefore formalized the problem of minimizing the \emph{risk} of polarization and disagreement (and more general objectives) with \emph{unknown} innate opinions. Let $G=(V,E,w)$ be an edge-weighted undirected graph and $\bm{s}\in [-1,1]^V$ an unknown, mean-centered innate-opinion vector. Let $\bm{z}^{(\infty)}$ be the limit of the opinion vector of the FJ model with degree-dependent susceptibility. Specifically, the authors considered a general objective, which they refer to as \emph{conflict}, encompassing the following four objectives: (i) internal conflict $\sum_{v\in V}(z^{(\infty)}_v-s_v)^2=\bm{s}^\top(I+L)^{-1}L^2(I+L)^{-1}\bm{s}$, (ii) disagreement $\sum_{\{u,v\}\in E}w_{uv}(z^{(\infty)}_u-z^{(\infty)}_v)^2=\bm{s}^\top(I+L)^{-1}L(I+L)^{-1}\bm{s}$, (iii) polarization $\|\bm{z}^{(\infty)}\|_2^2=\bm{s}^\top(I+L)^{-2}\bm{s}$, and (iv) polarization-disagreement index $\sum_{\{u,v\}\in E}w_{uv}(z^{(\infty)}_u-z^{(\infty)}_v)^2 + \|\bm{z}^{(\infty)}\|_2^2=\bm{s}^\top(I+L)^{-1}\bm{s}$.
Given that the innate-opinion vector $\bm{s}$ is unknown, they introduced two measures to quantify the risk of conflict: Average-case Conflict Risk (ACR) and Worst-case Conflict Risk (WCR). ACR is defined as the expected conflict, where the expectation is taken over innate opinions drawn uniformly at random from $\{-1,1\}^V$. This quantity can be represented as $\mathbb{E}[\bm{s}^\top M\bm{s}]=\mathrm{Tr}(M)$, where $M$ is an appropriate matrix depending on the objective at hand. WCR is a more robust measure, defined as the maximum (i.e., worst-case) conflict over all possible innate-opinion vectors in $\{-1,1\}^V$, which can be represented as $\max_{\bm{s}\in \{-1,1\}^V}\bm{s}^\top M\bm{s}$. Although computing this quantity is NP-hard, there exists an SDP-based $(2/\pi)$-approximation algorithm~\cite{Nesterov98}. Based on the above discussion, the authors formulated the following intervention problem:
\begin{problem}[Conflict Risk Minimization~\cite{Chen+18}]\label{prob:pd_conflict}
Given an edge-weighted undirected graph $G=(V,E,w)$ with edge weights in $[0,1]$, or equivalently its weighted adjacency matrix $W\in [0,1]^{V\times V}$, and a budget $b\in \mathbb{R}_{>0}$, the goal is to find a modified adjacency matrix $W'\in [0,1]^{V\times V}$ that minimizes ACR or WCR subject to $\|W-W'\|_1\leq 2b$. 
\end{problem}
For this problem, the authors proposed two algorithms, i.e., coordinate descent and conditional gradient descent, based on explicit gradient derivations. Their computational experiments showed that even a small amount of intervention quickly decreases the risk of conflict, in terms of both ACR and WCR. They also found that minimizing ACR often does not reduce WCR, whereas minimizing WCR, though computationally more challenging, effectively reduces ACR.

Later, \citet{Chaitanya+24} generalized the above problem with WCR by considering the $p$-th order controversy $\bm{s}^\top(I+L)^{-p}\bm{s}$ and by enlarging the set of possible innate-opinion vectors from $\{-1,1\}^V$ to $[-1,1]^V$. 
To design an effective algorithm, the authors derived an upper bound on the objective, based on the graph Laplacian, 
and showed that minimizing this upper bound is equivalent to maximizing the log-determinant of $I+L$, which is a convex program. 

\citet{Cinus+25} also studied polarization and disagreement minimization under uncertainty about innate opinions, but in a different setting. In their problem, interventions are limited to rebalancing the social feed on a social media platform, following \citet{Cinus+23}. The objective is to minimize the polarization, disagreement, or polarization-disagreement index for the unknown innate opinions, while being able to query the innate opinions of a limited number of nodes. Their problem is formulated as follows:
\begin{problem}[Polarization and Disagreement Minimization with Unknown Innate Opinions~\cite{Cinus+25}]\label{prob:pd_unknown}
Given an edge-weighted (un)directed graph $G=(V,E,w)$, or equivalently its weighted adjacency matrix $W\in \mathbb{R}_{\geq 0}^{V\times V}$, and a budget $b\in \mathbb{Z}_{\geq 0}$, the goal is to find a new adjacency matrix $W'$ from $W$ (as in Problem~\ref{problem:Cinus+23}) that minimizes the polarization, disagreement, or polarization-disagreement index, defined for the unknown innate-opinion vector $\bm{s}\in [-1,1]^V$, under the assumption that we can query the innate opinions of at most $b$ nodes. 
\end{problem}

For this problem, the authors developed a three-step approach: (1) select $b$ nodes and observe their innate opinions, (2) reconstruct the innate opinions of all nodes using the $b$ observed opinions and the network structure, and (3) optimize the objective function using the reconstructed opinions.
They first quantified how reconstruction error in innate opinions affects the final solution quality, thereby highlighting the importance of steps (1) and (2). Specifically, their analysis showed that the reconstruction error, together with the Lipschitz constant of the objective, bounds the optimization error.
For step (1), the authors proposed selecting the top-$b$ nodes according to centrality measures, including degree centrality, closeness centrality~\cite{Bavelas48}, and PageRank~\cite{Page+99}. For step (2), they considered three reconstruction methods based on label propagation, Graph Neural Networks, and graph signal processing, all motivated by the intuition that innate opinions exhibit homophily in real-world networks. Step (3) was implemented by applying existing polarization and disagreement minimization algorithms that assume full knowledge of innate opinions.
Their experiments showed that polarization and disagreement can be effectively minimized even with limited information about innate opinions.

Very recently, \citet{Cinus+25_2} studied a more realistic online setting in which innate opinions are unknown and must be learned through sequential observations. This novel setting, which naturally captures periodic interventions on social media platforms, was formulated as a regret-minimization problem, establishing a connection between algorithmic interventions and multi-armed bandit theory~\cite{Lattimore+20}. Specifically, the authors formulated the following problem: 
\begin{problem}[Online Minimization of Polarization and Disagreement~\cite{Cinus+25_2}]\label{prob:pd_bandit} 
Let $V$ be a set of nodes. Let $\mathcal{L} = \{L_1,L_2,\dots, L_K\}$ be an intervention space consisting of graph Laplacians over $V$ representing possible network structures. 
The online learning protocol proceeds in rounds $t = 1,\dots ,T$ as follows: (i) the learner selects an intervention $L^{(t)}\in \mathcal{L}$. The FJ dynamics converges to the limit and the learner incurs the loss, i.e., the polarization-disagreement index $f(L^{(t)}) \coloneqq \bm{s}^{\top}(I + L^{(t)})^{-1} \bm{s}$; 
(ii) the learner observes bandit feedback in the form of a noisy loss signal $f(L^{(t)}) + \eta_t$, where $\eta_t$ is a zero-mean random noise.  
The learner’s goal is to minimize the cumulative regret $R_T$,
defined as the difference between the cumulative loss of the chosen interventions and that of the best fixed intervention $L^* = \arg\min_{L \in \mathcal{L}} f(L)$, i.e.,
$R_T = \sum_{t=1}^T \left( f(L^{(t)}) - f(L^*) \right)$.

\end{problem}
For this sequential decision-making problem, the authors presented a two-stage algorithm based on low-rank matrix bandits~\cite{Lu+21,Kang+24}. The algorithm first performs subspace estimation to identify an underlying low-dimensional structure, and then applies a linear bandit algorithm in a reduced-dimensional representation induced by the estimated subspace. They showed that the algorithm achieves the cumulative regret of order $\sqrt{T}$ for any time horizon $T$. Empirical results showed that it outperforms a linear-bandit baseline in both cumulative regret and running time.

In contrast to the above studies on polarization and disagreement minimization, \citet{Gaitonde+20} initiated the study of polarization and disagreement \emph{maximization}, which pursues the opposite objective. This line of work helps clarify the potential risks of conflict and polarization in settings where an adversary seeks to amplify them. Specifically, the authors formulated the following problem:
\begin{problem}[Adversarial Perturbations of Opinion Dynamics~\cite{Gaitonde+20}]\label{prob:pd_perturbations}
Given an edge-weighted undirected graph $G=(V,E,w)$ and a budget $b \in \mathbb{R}_{>0}$, 
the goal is to find an innate-opinion vector $\bm{s}\in \mathbb{R}^V$ with $\|\bm{s}\|_2\leq b$ that maximizes the polarization, disagreement, or polarization-disagreement index of the limit of the expressed opinions under the FJ model with degree-dependent susceptibility. 
\end{problem}
Based on spectral techniques, the authors showed that, for different objectives, different spectral regimes of the graph limit the adversary’s capacity to sow conflict and polarization. Leveraging the strong connections between spectral and structural graph properties, they qualitatively described which networks are most vulnerable or resilient to such perturbations.
They also studied the algorithmic problem faced by a network defender, who seeks to mitigate these attacks by insulating nodes heterogeneously. By exploiting the geometry of the problem, they showed that the resulting optimization can be solved efficiently via convex programming. Finally, they generalized their results to a setting with two network structures, in which the opinion dynamics and the measurement of conflict and polarization are decoupled. For instance, this can arise when opinion dynamics unfold within an online community mediated by social media, while disagreement is measured over real-world connections. They characterized conditions on the relationship between the two graphs that determine how much additional power the adversary gains in this setting.

\citet{Chen+22_2} studied a variant in which the adversary is given an initial innate-opinion vector $\bm{s}\in [0,1]^V$ (not necessarily a zero vector) and a budget $b\in \mathbb{Z}_{>0}$, and is asked to find a modified innate-opinion vector $\bm{s}'\in [0,1]^V$ with $\|\bm{s}-\bm{s}'\|_0\leq b$ maximizing one of the objectives. The authors first highlighted the combinatorial nature of the problem: regardless of which set of nodes the adversary selects, the optimal way to modify their innate opinions is to push them to one of the two extremes, i.e., $0$ or $1$, which follows from the convexity of the objective functions. They then showed that the increase in the polarization-disagreement index is linear in $b$, while the increase in disagreement is only bounded by $b\Delta$, where $\Delta$ is the maximum degree of the graph. Their computational experiments demonstrated that even simple heuristic methods can achieve objective increases proportional to $b$, highlighting the practical risks of attacks that amplify polarization and disagreement.

\citet{Tu+23} further investigated a variant in which the adversary has limited information: the adversary observes only the network structure, without access to individuals’ innate opinions, and seeks to increase disagreement by radicalizing a subset of users. A central contribution is a formal connection between the full- and limited-information settings. Under mild assumptions, the authors showed that radicalizing users who are structurally influential for network disagreement yields an $O(1)$-approximation to the optimal full-information adversary.
From the computational complexity perspective, they proved that maximizing disagreement in the FJ model with degree-dependent susceptibility, with the interventions on innate opinions, is NP-hard. 
They further showed that the problem reduces to a cardinality-constrained MaxCut problem on graphs with both positive and negative edge weights, and presented the first $O(1)$-approximation algorithm for this problem, 
based on an SDP relaxation and randomized hyperplane rounding~\cite{goemans1995improved}. Extensive experiments showed that topology-only adversarial strategies outperform simple baselines and achieve performance comparable to full-information algorithms.
Recently, \citet{Zareer+25} studied a counterpart in which the network structure is not visible initially. They devised an algorithm based on Double Deep Q-learning.

\smallskip
\noindent
\textbf{Intervention-Related Insights.}
\citet{Chitra+20} studied how algorithmic filtering on social media platforms can drive polarization. They extended the FJ model with degree-dependent susceptibility by introducing an external actor representing a platform provider that dynamically adjusts the strength of user interactions to increase user engagement by reducing disagreement. Using this model, they empirically showed on Reddit and Twitter datasets that even small changes by platform providers can increase user polarization.
Similarly, \citet{Bhalla+23,Bhalla+25} studied how local edge dynamics, combined with the FJ model, drive polarization over time. They showed that confirmation bias together with friend-of-friend recommendations can erode cross-group connectivity, thereby increasing polarization.
On the other hand, \citet{Wang+23} showed that recommender-driven objectives (namely, maximizing relevance) are not inherently incompatible with interventions aimed at minimizing polarization and disagreement. 
\citet{Tu+22} studied the impact of viral information on the opinions in networks. They introduced a model in which an information spreads through the network (as in the IC model~\cite{Kempe+03,Kempe+15}) and may change nodes' innate opinions. They showed that if innate opinions can only be increased, this has a much smaller effect on the polarization and disagreement in the network than when the innate opinions can change in both directions. 
Recently, \citet{Shirzadi+25_2} studied how users’ stubbornness affects polarization and disagreement in the FJ model. They showed that increasing homogeneous stubbornness always amplifies the polarization-disagreement index, whereas in the heterogeneous case, increasing the stubbornness of neutral individuals can reduce the index.
\citet{Biondi+23} collected a variety of polarization and disagreement indices for the FJ model and its variants, and derived conditions under which opinion dynamics can amplify polarization or disagreement, in the sense that the expressed opinions exhibit greater polarization or disagreement than the innate opinions.

Very recently, \citet{Jalan+26} generalized the threat model by considering multiple, decentralized strategic actors with potentially conflicting goals under the FJ model.
In their setting, a subset of strategic actors $S \subseteq V$ manipulates the network by replacing the true innate-opinion vector $\bm{s} \in \mathbb{R}^V$ with a fictitious innate-opinion vector $\bm{s}' \in \mathbb{R}^V$. Specifically, each strategic actor $v \in S$ unilaterally reports a manipulated innate opinion $s'_v \neq s_v$, while honest agents $u \in V \setminus S$ report their true innate opinions $s'_u = s_u$. The authors formalized this manipulation as a meta-game where each actor $v \in S$ strategically chooses $s'_v$ to minimize their individual cost defined by \citet{Bindel+15,Bindel+11} (see the beginning of Section~\ref{subsec:pd_disagreement}) evaluated at the resulting equilibrium $\bm{z}^{(\infty)}(\bm{s}')$, and they characterized the pure strategy Nash equilibrium of this game.
To quantify the resulting societal harm, they introduced the Price of Misreporting (PoM)---defined analogously to the price of anarchy as the ratio of the social cost at the manipulated limit of the expressed opinions to that at the truthful equilibrium.
Empirically, the authors demonstrated that multiple adversaries drastically inflate both polarization and disagreement, driving up the PoM. 
To defend against this, they proposed a diagnostic intervention: when $|S|$ is sufficiently small, their polynomial-time algorithm uses robust regression on the observed expressed opinions to recover the exact set of strategic actors $S$.

\section{Optimization of Other Objectives}\label{sec:others}

Here we review algorithmic-intervention studies that consider objectives beyond overall opinion and polarization and disagreement. For clarity, we still organize the objectives into three groups: (i)~objectives related to overall opinion, (ii)~objectives related to polarization and disagreement, and (iii)~additional objectives, to highlight the underlying intuitions.
Table~\ref{tab:summary_others} provides an overview of the literature. 

\begin{table}[t]
\caption{Summary of the literature on optimizing objectives beyond overall opinion and polarization and disagreement. 
Same notation and categorization rule as in Tables~\ref{tab:overall_summary} and~\ref{tab:summary_pd}.}
\label{tab:summary_others}
\resizebox{\textwidth}{!}{
\begin{tabular}{cccccccc}
\toprule
Objective & Intervention & Reference & Graph & Opinion-dynamics model & Main results \\
\midrule
Generalized Overall Opinion & Network structure & \citet{amelkin2019fighting} &W--D & Not assumed & Heuristic alg. \\
 & Leader selection & \citet{Saha+23} &W--D & LF FJ (gen) with multiple topics & Approx. alg. \\
 & & \citet{Liu+25_2} &U--U & Multifaceted opinion evolution & Heuristic alg. \\
Cumulative Overall Opinion & Leader selection & \citet{Fardad+13} &W--D & LF DeGroot & Heuristic alg. \\
 & & \citet{Mai+19} &W--D & LF DeGroot variant & Approx. alg. \\
 &Innate opinions & \citet{Cianfanelli+25} &W--D & FJ (gen) & MILP formulations \\
Median Opinion & Susceptibility to persuasion & \citet{Ristache+25} &W--U & FJ (gen) & Heuristic alg. \\
\midrule
Diversity of Opinions & Leader selection & \citet{Mackin+19} & U--U & LF DeGroot & Analytical solutions \\
Perception Gap & Network structure & \citet{Gehlot+25} &U--U & Not assumed & Heuristic alg. \\
\midrule
Convergence Speed & Leader selection & \citet{Bokar+15} & W--D & LF DeGroot & Approx./Heuristic alg. \\
 & & \citet{Zhou+25} &U--U & LF consensus & Heuristic alg. \\
 & Network structure & \citet{Zhou+23_2} &U--U & LF consensus & Approx. alg. \\
Social Power & Network structure & \citet{Wang+25_2} & W--D & FJ (gen) & Approx. alg. \\
Disparity & Innate opinions (and others) & \citet{Papachristou+25} &W--D & DeGroot/FJ (deg) & Exact/Approx. alg. \\
\bottomrule
\end{tabular}
}
\end{table}

\subsection{Objectives Related to Overall Opinion}\label{subsec:others_overall}

\noindent
\textbf{Generalized Overall Opinion.} \citet{amelkin2019fighting} introduced a node-weighted variant of the overall opinion. For an opinion vector $\bm{z}\in [0,1]^V$, they defined it as $\bm{\pi}^\top\bm{z}$, where $\bm{\pi}\in \mathbb{R}^V$ is the vector of nodes' eigenvector centrality scores~\cite{Bonacich72}. In this variant, opinions of more central nodes receive higher weight.
The authors formulated the problem of recovering the weighted overall opinion after an attack by a malicious adversary, where the adversary manipulates individuals' opinions and one seeks to restore the weighted overall opinion by adding a limited number of edges. They proved NP-hardness and presented a nearly linear-time heuristic based on a perturbation analysis. 
\citet{Saha+23} introduced voting-based measures for scenarios with competing campaigns, which aggregate individuals' opinions under different voting settings, and studied the corresponding opinion maximization problems.
Closely related to this, \citet{Liu+25_2} considered a variant of weighted overall opinion, where each node’s opinion is weighted by its communicability score~\cite{Estrada+08} within the cluster to which it belongs, and studied the corresponding opinion maximization problem via interventions on innate opinions. 

\smallskip
\noindent
\textbf{Cumulative Overall Opinion.}
\citet{Fardad+13} studied convergence-speed optimization under the LF DeGroot model. In their problem, given an edge-weighted directed graph $G=(V,E,w)$, an initial opinion vector $\bm{x}^{(0)}=\bm{1}$, and a budget $b\in \mathbb{Z}_{>0}$, the goal is to find a leader set $L\subseteq V$ with $|L|\leq b$, where leaders hold opinion $0$, that minimizes the cumulative overall opinion $\sum_{t=0}^{\infty}\sum_{v\in V\setminus L} x^{(t)}_F(L)_v$ of the LF DeGroot model (see Section~\ref{subsec:DeGroot}). For this problem, the authors presented an efficient heuristic based on a continuous relaxation and coordinate descent. They also formulated a variant in which the leader set is fixed and the intervention is instead performed on the network structure.
Later, \citet{Mai+19} studied a generalized setting in which leaders are not completely stubborn and the initial opinion vector can be arbitrary.
Recently, \citet{Cianfanelli+25} studied the problem of maximizing the cumulative overall opinion through interventions on innate opinions under uncertainty in the network structure. They avoided relying on exact network topologies by assuming a time-varying random network model based solely on macroscopic node statistics.

\smallskip
\noindent
\textbf{Median Opinion.} \citet{Ristache+25} proposed focusing on the median opinion, motivated by settings in which elections need to be protected from improper influence. In their problem, given an edge-weighted undirected graph $G=(V,E,w)$, an innate-opinion vector $\bm{s}\in [0,1]^V$, a susceptibility vector $\bm{\alpha}\in [0,1]^V$, and a budget $b\in \mathbb{R}_{>0}$, the goal is to find a modified susceptibility vector $\bm{\alpha}'\in [0,1]^V$ that maximizes the median opinion at the FJ equilibrium, subject to $\|\bm{\alpha}-\bm{\alpha}'\|_p\leq b$ for some $p\geq 0$. The authors proved that the problem is NP-hard and inapproximable within any constant factor. They then presented three efficient heuristics. The first two take a continuous approach: one employs gradient descent with Huber’s estimator to approximate the median, and the other uses a sigmoid threshold influence function. The third uses a combinatorial greedy strategy for targeted interventions. Their empirical results showed that even small interventions can significantly shift the median, thereby potentially swaying election outcomes.

\subsection{Objectives Related to Polarization and Disagreement}\label{subsec:others_pd}

\noindent
\textbf{Diversity of Opinions.}
\citet{Mackin+19} introduced two measures for quantifying the diversity of opinions, namely the Simpson Opinion Diversity Index and the Shannon Opinion Diversity Index, by adapting the Simpson Diversity Index~\cite{Simpson49} and the Shannon Index~\cite{Shannon48}, respectively. 
Based on these indices, the authors formulated the problem of maximizing opinion diversity by locating a leader with opinion $1$ in a graph that already contains a leader with opinion $0$ under the LF DeGroot model. 
They provided analytical solutions for some classes of graphs. 

\smallskip
\noindent
\textbf{Perception Gap.} Recently, \citet{Gehlot+25} introduced a measure of perception gap, quantifying how much individuals’ average views within their social groups deviate from the average opinion of the entire network. Let $G=(V,E)$ be an unweighted undirected graph with an opinion vector $\bm{z}\in \mathbb{R}^V$. The authors defined the perception gap to be $\sum_{v\in V}\left(\frac{\sum_{u\in N(v)}z_u}{|N(v)|}-\frac{\sum_{u\in V}z_u}{n}\right)^2$. They formulated the problem of minimizing perception gap via interventions by adding a limited number of edges. They proved an inapproximability result and presented efficient heuristics.

\subsection{Additional Objectives}\label{subsec:others_additional}

\noindent
\textbf{Convergence Speed.}
\citet{Bokar+15} studied an objective related to the convergence speed to the equilibrium under the LF DeGroot model. The authors aimed to minimize the Perron--Frobenius eigenvalue of the submatrix (corresponding to followers) of the (row-)random-walk matrix $P$ defined from the weighted adjacency matrix of $G$, motivated by the fact that this eigenvalue governs the convergence rate to consensus. On the other hand, \citet{Zhou+23_2} studied maximizing the smallest eigenvalue via edge-addition interventions.

\smallskip
\noindent
\textbf{Social Power.} 
Recently, \citet{Wang+25_2} addressed the problem of maximizing the social power of an external influencer under the general FJ model.
Recall that the equilibrium opinion vector is given by 
$\bm{z}^{(\infty)} = (I - \Lambda P)^{-1} (I-\Lambda) \bm{s}$.
The social power of an individual is then defined as the corresponding column sum of the matrix $(I - \Lambda P)^{-1} (I-\Lambda)$, which represents the overall influence of that individual's innate opinion on the equilibrium opinions~\cite{Jia+15,Tian+22}.
In their optimization problem, given an edge-weighted directed graph $G=(V,E,w)$ and a budget $b\in \mathbb{Z}_{>0}$, the goal is to add $b$ edges from nodes in the graph to the external influencer (modeled as a singleton) so as to maximize the influencer's social power.
The authors presented a $(1-1/\mathrm{e})$-approximation algorithm via submodularity of the objective. 

\smallskip
\noindent
\textbf{Disparity.}
Recently, \citet{Papachristou+25} introduced the disparity measure, which quantifies differences in the power of influence across groups in opinion dynamics. Let $G=(V,E,w)$ be an edge-weighted directed graph, where $V$ is partitioned into $A$ and $B=V\setminus A$. Let $\bm{s}\in \mathbb{R}^V$ be the initial opinion vector in the DeGroot model or the innate-opinion vector in the FJ model with degree-dependent susceptibility, where $\|\bm{s}\|_2=1$. Let $\bm{s}_A\in \mathbb{R}^V$ be the modified innate-opinion vector obtained by keeping all innate opinions of nodes in $A$ but setting all innate opinions of nodes in $B$ to $0$. The modified innate-opinion vector $\bm{s}_B\in \mathbb{R}^V$ is defined in a similar manner. Let $\bm{z}_A\in \mathbb{R}^V$ and $\bm{z}_B\in \mathbb{R}^V$ be the equilibrium opinions induced by $\bm{s}_A$ and $\bm{s}_B$, respectively, under the DeGroot model or the FJ model. Then the disparity is defined as $\|\bm{z}_A-\bm{z}_B\|_2$, capturing how different the equilibrium outcomes are when only one group’s opinions are retained at a time.
The authors formulated and studied a variety of disparity minimization and maximization problems via interventions in initial or innate opinions, network structure, and partition selection, under the DeGroot model or the FJ model. For instance, they showed that in the DeGroot model, minimizing disparity via interventions in initial opinions or network structure is solvable in polynomial time, whereas finding a partition that minimizes disparity is NP-hard.

\section{Conclusion and Future Directions}\label{sec:conclusion}

In this survey, we have provided a structured overview of the rapidly growing literature on algorithmic interventions in opinion dynamics. We organized prior work primarily by the objective it seeks to optimize, including overall opinion (Section~\ref{sec:overall}), polarization and disagreement (Section~\ref{sec:pd}), and other quantities (Section~\ref{sec:others}). 
For each category, we reviewed the underlying optimization formulations and representative algorithms, covering both classic and recent results in a mathematically grounded manner.
Beyond explicit intervention models and algorithms, we also summarized intervention-relevant insights---studies that do not directly propose an intervention mechanism in opinion dynamics but offer useful guidance for intervention design. 

Below, we outline concrete future directions that emerge from this survey.

\smallskip
\noindent
\textbf{Interventions beyond the DeGroot and FJ Models.} As discussed in Sections~\ref{sec:overall}--\ref{sec:others}, most existing approaches to algorithmic intervention in opinion dynamics are designed around the DeGroot and FJ models. However, many other opinion-dynamics models have been proposed (see Section~\ref{subsec:models_others}), and interventions tailored to these models remain largely underexplored, potentially limiting applicability across real-world settings. Accordingly, a natural and promising direction is to study algorithmic interventions under broader opinion-dynamics models.

\smallskip
\noindent
\textbf{Interventions with Limited Information and Uncertainty.} Most existing approaches to algorithmic intervention in opinion dynamics assume full access to the data specifying the underlying model (e.g., the network structure and the initial or innate opinions). In practice, however, this assumption often fails due to privacy constraints, incomplete observations, and measurement noise. An early attempt to address this issue was made by \citet{Chen+18}, followed by \citet{Chaitanya+24}, \citet{Cinus+25,Cinus+25_2}, and \citet{Cianfanelli+25}. 
Nevertheless, these works focused only on uncertainty in innate opinions under the FJ model and did not consider other sources of uncertainty, 
with the exception of \citet{Cianfanelli+25}, who studied the problem of maximizing the cumulative overall opinion under uncertainty in the network structure.
A promising direction is to study algorithmic interventions for a variety of objectives under broader uncertainty models.

\smallskip
\noindent
\textbf{Investigating Synergies with Graph Neural Networks (GNNs).}
While only a few studies have made this connection explicit, there may be fruitful synergies between algorithmic interventions in opinion dynamics and GNNs~\cite{Scarselli+08,Wu+21}.
Most existing intervention approaches rely on analytically tractable dynamics and equilibrium characterizations; it remains open how learning-based dynamical models could contribute in more realistic settings.
When opinion dynamics are unknown, partially observed, or nonlinear, GNNs may serve as surrogate dynamical models learned from trajectories, enabling the design of intervention algorithms that operate beyond closed-form equilibria. 
Recent work has begun to bridge opinion dynamics and GNNs by learning surrogate dynamics from trajectories while preserving equilibrium-related phenomena~\cite{Li+25}. 
Conversely, intervention primitives suggest useful perspectives for GNN design and training.
For instance, oversmoothing can be viewed as an analogue of consensus formation, suggesting that disagreement-maximization objectives may inspire architectural modules or regularizers that mitigate feature collapse.
More broadly, common intervention primitives such as \emph{rewiring} and \emph{leader selection} naturally translate to learning utilities:
rewiring corresponds to task-aware graph rewiring (especially under heterophily), and leader selection corresponds to selecting informative nodes to query in graph active learning.
Overall, tighter integration of these perspectives could yield interventions that remain effective under uncertainty and GNNs that are more robust and interpretable.

\smallskip
\noindent
\textbf{Bridging Theory and Practice.}
While the line of work covered in this survey has developed a rich mathematical foundation, its practical effectiveness in real-world deployments remains poorly understood. Empirical evaluation is typically limited to stylized experiments on existing datasets and synthetic simulations; in contrast, studies that assess interventions in live social media platforms remain scarce.
A key obstacle to bridging this gap is that realistic intervention experiments are often infeasible without direct access to platform infrastructure, limiting researchers’ ability to manipulate product features and run controlled online studies at scale.
One promising direction is to develop controllable, high-fidelity social-media testbeds that emulate platform dynamics (e.g., feed ranking and recommendation).
Recent work has begun to pursue this approach, for example by building research platforms that use AI agents to replicate live social-media interactions for controlled trials~\cite{Neubrander+26}.
Such simulators can complement stylized evaluation by enabling repeatable, high-coverage experiments that stress-test intervention mechanisms before deployment.

\section*{Acknowledgments}
This work is partially supported by Japan Science
and Technology Agency (JST) Strategic Basic Research
Programs PRESTO ``R\&D Process Innovation by AI and
Robotics: Technical Foundations and Practical Applications''
grant number JPMJPR24T2
and Vienna Science and Technology Fund (WWTF) [Grant ID: 10.47379/VRG23013].

\bibliographystyle{ACM-Reference-Format}
\bibliography{main}

\end{document}